\documentclass[superscriptaddress,amssymb,twocolumn,pre]{revtex4-2}

\usepackage{natbib,graphicx,dcolumn,bm,amsmath,color,bm,subfigure}

\newcommand{\eq}[1]{Eq.~(\ref{#1})}

\newcommand{\fig}[1]{Fig.~\ref{#1}}

\definecolor{amber}{rgb}{1.0, 0.75, 0.0}

\newcommand{\eeq}{ \end{equation} }
\newcommand{\beq}{ \begin{equation} }

\newcommand{\eea}{ \end{align} }
\newcommand{\bea}{ \begin{align} }

\newcommand{\oma}{{\boldsymbol \omega}}
\newcommand{\omb}{{\boldsymbol \omega}^{\prime} }

\newcommand{\bor}{ {\bf r} }
\newcommand{\borp}{ {\bf r}^{\prime}}

\newcommand{\borpe}{ {\bf r}_{ \perp} }
\newcommand{\borpeb}{ {\bf r}_{ \perp}^{\prime}  }

\begin{document}

\title{Elastic response of colloidal smectics: insights from microscopic theory}

\author{H. H. Wensink}
\affiliation{Laboratoire de Physique des Solides - UMR 8502, CNRS,  Universit\'{e} Paris-Saclay, 91405 Orsay, France}
\email{rik.wensink@universite-paris-saclay.fr}
\author{E. Grelet}
\affiliation{Centre de Recherche Paul Pascal - UMR 5031,  CNRS, Universit\'{e} de Bordeaux, 33600 Pessac, France}

\begin{abstract}

Elongated colloidal rods at sufficient packing conditions are known to form stable lamellar or smectic phases. Using a simplified volume-exclusion model, we propose a generic equation-of-state for hard-rod smectics that is robust against simulation results and is independent of the rod aspect ratio. We then extend our theory by  exploring  the elastic properties of a hard-rod smectic, including the layer compressibility ($B$) and bending modulus ($K_{1}$). By introducing weak backbone flexibility we are able to compare our predictions with experimental results on smectics of filamentous virus rods ({\em fd}) and find quantitative agreement between the smectic layer spacing, the out-of-plane fluctuation strength, as well as the smectic penetration length  $\lambda = \sqrt{K_{1}/B}$. We demonstrate that the layer bending modulus is dominated by director splay and depends sensitively on lamellar out-of-plane fluctuations that we account for on the single-rod level. We find that the ratio between  the smectic penetration length and the lamellar spacing is about two orders of magnitude smaller than typical values reported for thermotropic smectics.  We attribute this to the fact that colloidal smectics are considerably softer in terms of layer compression than their thermotropic counterparts while the cost of layer bending is of comparable magnitude.

\end{abstract}

\date{\today}

\maketitle 

\section{Introduction}

Non-spherical colloidal particles or elongated molecules form liquid crystal mesophases with properties interpolating between those of a fluid, such as liquid-like diffusivity, and those of a crystal such as long-ranged periodic order \cite{de1993physics}. Because of these hybrid properties, liquid crystals find widespread use in technological applications and knowledge of their phase behavior and mechanical response is of key importance in controlling and optimizing material properties composed of strongly non-spherical moieties \cite{de1980physical}. Concepts of liquid crystal physics are also instructive to understand processes in the living cell \cite{doostmohammadi2021physics,jewell2011living} and identifying structures in biological matter \cite{mitov2017cholesteric}. 

Nematic phases are the simplest of such structures and are characterized by long-ranged orientational order of the particles,
while lacking positional order in any direction. Moving one level up in order we find Smectic A-type order  that combines long-ranged stacking order along the main direction of particle alignment while retaining fluidity in the plane transverse to the layer normal. As such these lamellar structures can be interpreted in terms of a combination of unidirectional order and bidimensional fluidity which endows these materials with specific material properties owing to their distinct membrane-like character \cite{helfrich1975out,helfrich1978steric,dejeu2003structure}.    

As for the colloidal domain, phase transition amongst nematic, smectic and other phases that may appear upon variation of particle density have been the subject of intense theoretical study (the reader is referred to a number of excellent review papers on this 
topic \cite{vroege1992phase,allen1993hard,singh2000phase,mederos_overview2014}). Minimal models that account for the stability nematic phases \cite{onsager1949effects} and those with lower symmetry such as smectic \cite{frenkel1988thermodynamic} and columnar organizations \cite{frenkel1989invited,veerman1992phase} are based on simple uniaxial shapes (such as cylinders) that experience hard or steeply repulsive interactions.

For smectic materials, theoretical efforts have been aimed mostly at analyzing mechanical instabilities based on continuum theory where the underlying microscopic features of the particles  are ignored. On the other hand, for rod-shaped colloids microscopic theories based on density functional concepts and simulations have primarily addressed the role of smectic order in  the overall phase diagram of non-spherical colloids \cite{frenkel1988thermodynamic,mcgrother1996re,bolhuis1997tracing,King2023}, subtle intralayer order \cite{van1995transverse} and, more recently,  also their diffusive properties \cite{grelet2008dynamical,patti2010collective, Chiappini2020}.

In this work, we aim to take a more in-depth look at colloid-based smectic liquid crystals taking a hard-core cylinders as benchmark model. We wish to not only address thermodynamic properties  such as the equation-of-state  but also various mesoscopic properties including the layer spacing, thermal corrugation of the smectic layers, and the elastic properties related to layer compression and layer bending. The latter are routinely described by the layer compression and bending moduli, $B$ and $K_{1}$, respectively \cite{de1993physics}.  Our findings are relevant for a variety of colloidal smectic systems whose lamellar structure is stabilized primarily (but not exclusively) by hard-core volume exclusion such as TMV \cite{wen1989observation,meyer1990ordered}, ${\rm TiO_{2}}$  \cite{hosseini2020smectic}, $\beta$-FeOOH \cite{Maeda2003}, goethite \cite{vroege2006smectic}, CdSe \cite{querner2008millimeter} and gold nanorods \cite{hamon2012three}. All smectic structures observed thus far were of the SmA or SmB type. While a SmA phase retains full intralayer fluidity, the SmB phase is characterized by long-ranged positional order of some kind within each layer.  Examples of SmC type layering, in which the colloidal rods are tilted with respect to the layer normal, are scarce. Observations of ordering of this kind have been reported for silica rods subjected to gravitational compression  \cite{kuijk2012phase}.

A number of recent experimental studies have been devoted to classifying topological defects in strongly confined colloidal smectics \cite{monderkamp2021topology,wittmann2021particle,wittmann2023smectic}. The symmetry and spatial extent of these defects depend strongly on the elastic properties of the smectic material which we intend to address here from a microscopic theoretical viewpoint.
Further inspiration for undertaking a study of smectic elastic moduli stems from the  emergence of advanced continuum theories for smectics \cite{xia2021structural, paget2022smectic,paget2023complex} that rely on the elastic moduli to predict the response of layered materials to geometric frustration and other external stimuli.

Our approach is based on a two-way extension of the commonly employed cell theory for positionally ordered liquid crystals \cite{hentschke1989equation,graf1997cell,graf1999phase,wensink2004equation,peters2020algebraic}.  The first is a simple scaling description for a single lamellar fluid composed of near-parallel rods by decoupling the effects of orientational fluctuations from the parallel-core contribution. We demonstrate that the scaling approach gives a correct rendering of the onset of smectic and columnar order from a nematic melt for rod- and disc-shaped colloids alike.  The second modification of the original cell-approach involves taking into account the out-of-plane fluctuations at the single-rod level and highlight their subtle role in determining the elastic properties of a smectic material, mostly in the layer bending modulus $K_{1}$. By suitably renormalizing quantities in terms of the individual rod dimensions we show  that the equation-of-state and the  elastic moduli of a hard-rod smectic can be rendered {\em independent} of the rod aspect ratio.  In particular, we demonstrate that the smectic penetration length $\lambda = \sqrt{K_{1}/B}$, which should be be interpreted as the intrinsic length scale over which elastic distortions such as edge and screw-dislocations relax in a lamellar system, does not primarily depend on the rod shape but only on the overall packing fraction of the smectic material.

We subsequently discuss the role of backbone flexibility of the rods which is relevant for filamentous virus rods and compare our predictions with experimental measurements on smectic phases of {\em fd} suspensions \cite{dogic1997smectic,grelet2008dynamical}. We find good agreement for the smectic layer spacing as a function of rod concentration as well for the typical smectic potential that controls the single-rod diffusivity \cite{lettinga2007self, pouget2011dynamics,Repula2019}. Last but not least, we predict that the smectic penetration $\lambda=\sqrt{K_{1}/B} \approx 0.02L$ is only a fraction of the rod length $L$ and the smectic layer distance, in excellent agreement with measured values for {\em fd}. This illustrates that a hard-rod smectic has a different mechanical response than the more commonly explored molecular-based lamellar systems, such as lipid membranes \cite{ollinger2005lipid} and thermotropic smectics where the smectic penetration length was found to be comparable to the lamellar periodicity \cite{de1993physics,Birecki1976,Fromm1987,Lewis1988}. We demonstrate that  this discrepancy is chiefly due to colloidal smectics having a considerably smaller layer compressiblity modulus than their thermotropic counterparts.

The remainder of this paper is organized as follows. In Section II and III we discuss a  density-functional theory for a lamellar fluid by allowing for Gaussian out-of-plane fluctuations at the single particle level. In Section IV we discuss the role of  orientational fluctuations within the lamellae and we address out-of-plane positional fluctuations in Section V. Section VI is devoted to comparing the equation-of-state of the a smectic bulk phase with simulation results.  We then move on to discussing the elastic properties in Sections VII to IX. A comparison with experimental smectics of both lyotropic and thermotropic nature is discussed in Section X.   Some concluding remarks are formulated in Section XI.  In the  Appendix we test our simplified description of the excluded volume interactions  between hard colloidal bodies within a comprehensive bifurcation analysis to identify bulk freezing transitions from a 3D nematic fluids composed of rod or disc-shaped colloids as well as freezing a monolayer fluid of rods and compare our predictions against simulation results.

\section{Density functional theory for a monolayer rod fluid}

\begin{figure*}
	\includegraphics[width = 1.7\columnwidth]{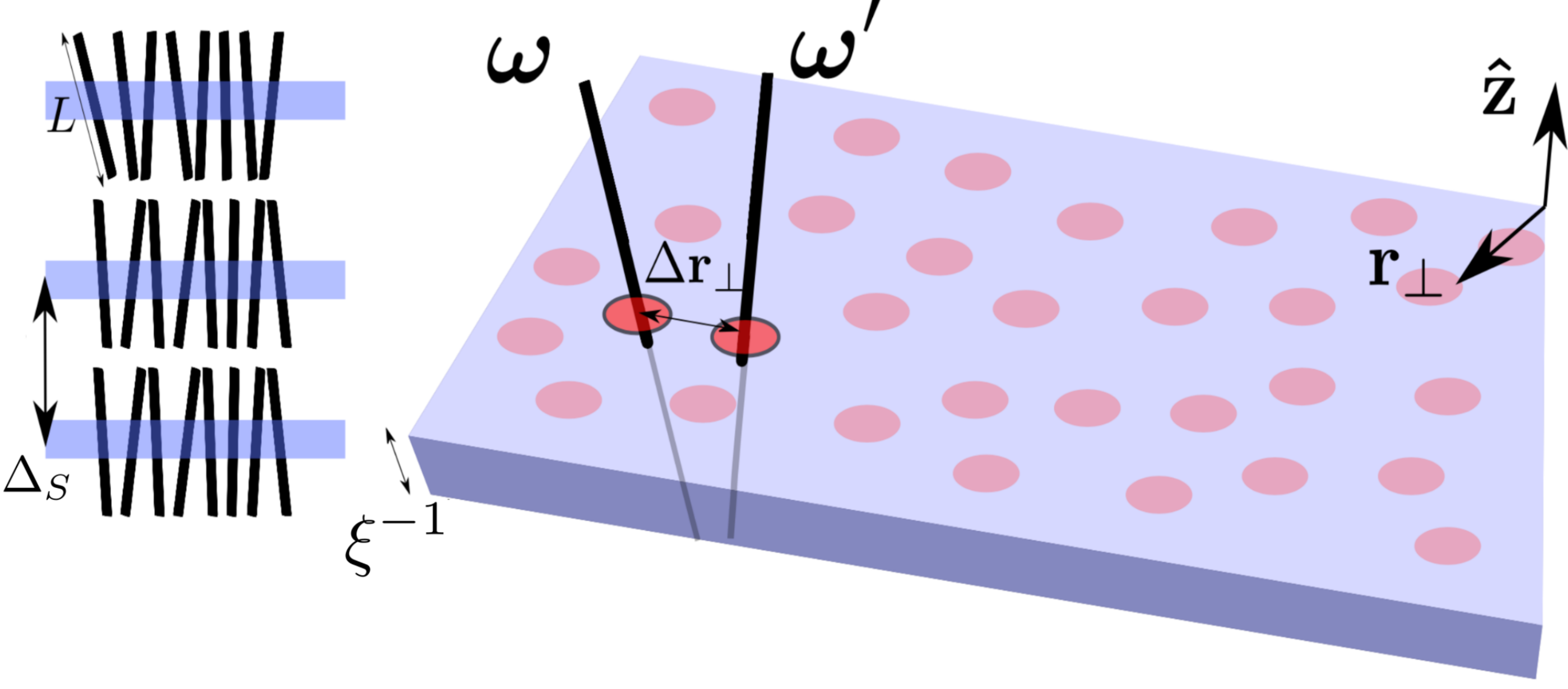}
	\caption{(left) Sketch of a smectic-A phase of colloidal rods of length $L$ with layer spacing $\Delta_{S}$.  (right) Illustration of the lamellar fluid model. Each lamella is composed of strongly aligned rods  with centres-of-mass (red dots) roughly confined to a narrow slab of width $\xi^{-1}$ as indicated in blue.   }
	\label{spacing}
\end{figure*}

Let us consider a monolayer fluid consisting of a collection of $N$ slender rods of length $L$ and diameter $D \ll L$ each with their centre-of-mass confined to a 2D $xy$-plane with area $A$. The rods are densely packed at a concentration $\rho_{\perp} = N/A \sim D^{-2}$ with their main axes aligned  along the out-of-plane $z$-direction (Fig. \ref{spacing}), while each rod experiences weak fluctuations about its average orientation.

The pair interaction $u_{r}$ between two rods with long-axis unit vectors $\oma$ and $\omb$ and centre-of-mass distance $\Delta \bor$ follows from short-range steric forces
\beq
\beta u_{r} (\Delta \bor; \oma, \omb) =
\begin{cases}
\infty & \Delta \bor < \sigma (\Delta {\bf \hat{r}}; \oma, \omb) \\
0 & \text{otherwise}, 
\end{cases}
\label{urod}
\eeq
where $\beta^{-1} = k_{B}T$ denotes the thermal energy and $\sigma$ is the contact distance between the hard cores of the rods which depends non-trivially on the centre-of-mass distance unit vector $\Delta {\bf \hat{r}} = \Delta \bor/\Delta r$ (for hard spheres $\sigma$ simply coincides with the sphere diameter).

Next, we apply a second-virial approximation by considering interactions between rod pairs alone. Later on we will heuristically account for higher-order virial terms through a simple density rescaling. The grand potential $\Omega$ for the lamellar fluid can be formally written as follows (dropping the orientational degrees of freedom for notational convenience)
\begin{align}
\beta \Omega[\rho] = &  \int d \bor \rho(\bor)(\ln  \nu \rho(\bor) -1 - \beta \mu) \nonumber \\ 
&- \frac{1}{2} \int d \bor \rho(\bor) \int d \borp \rho(\borp) \Phi (\Delta \bor ),
\label{fex}
\end{align}
with $\Delta \bor = \bor - \bor^{\prime}$ and $\Phi$ the Mayer function that features the pair potential
\beq
\Phi (\Delta \bor) = e^{-\beta u_{r}(\Delta \bor) }-1.
\eeq
Most importantly, $\rho(\bor) = \rho(\bor; \oma)$ denotes the one-body rod density while the chemical potential $\mu$ ensures its  proper normalization,  $\int d\bor \int d \oma  \rho(\bor ; \oma) = N$. Finally, $\nu$ is the effective thermal volume of each rod which also contains contributions from its rotational momenta. For spherical particles, $\nu=\Lambda^3$ with $\Lambda$ the thermal de Broglie wavelength.

The first term in the free energy  \eq{fex} describes the ideal translational and rotational entropy of each rod while the second term relates to the excess free energy that encodes rod-rod correlations at the level of pair interactions.
Inside the lamella  rod positions are strongly confined around $z=0$ while rods are free to move within the lamellar plane ($xy$) that we parameterize by the 2D vector $\bor_{\perp}$.  The easiest way to constrain the rods to remain in a monolayer configuration  would be to introduce an external harmonic potential, e.g. $U_{\rm lam}(z) \propto  U_{0} z^{2} $ that would penalize  out-of-plane excursions with $U_{\rm lam}$ the typical (entropic) energy barrier associated with rods hopping from one layer to the next. 
Once the smectic potential has been specified one  needs to resolve the one-body density $\rho(\bor)$ associated with these smectic potentials keeping in mind that this object depends on three positional variables and on two orientational ones.   

In this paper we adopt a more manageable strategy by {\em constraining} the one-body density to take on a simple factorized form
\beq
\rho (\bor; \oma   ) = G(z) \rho(\bor_{\perp}; \oma). 
\label{rho3d}
\eeq
We invoke a cylindrical coordinate frame $ \bor =  \borpe + z {\bf \hat{z}}$ with the planar coordinate denoted by the subscript $\perp$ and the out-of-plane one by $z$. Then, $G$ denotes a Gaussian  fluctuation ($z \in [-\infty, \infty]$)
\beq
G(z)= \left ( \frac{\xi}{\pi} \right )^{1/2} \exp (- \xi z^{2} ),
\label{geezer}
\eeq
where  $(2 \xi)^{-1}  = \langle  z^{2} \rangle $ quantifies the mean-squared value for the out-of-plane excursions which are expected to be very small compared to the rod length, i.e. $\langle z^{2} \rangle \ll L^{2}$. Then, $\lim G_{\xi \rightarrow \infty}$ describes the limiting case of perfectly confined rods. The above parameterization implies that the in-plane positional correlations are coupled to the rod orientations while both are assumed to be unaffected by  the out-of-plane rod fluctuations. 

Inserting \eq{rho3d} into the grand potential and minimizing with respect to the unknown planar density $\rho(\bor_{\perp}; \oma)$ we obtain the Euler-Lagrange (EL) equation (again dropping all dependencies on $\oma$ to prevent cluttered notation)
\beq
\ln \nu_{2} \rho(\borpe) - \int d \borpeb \rho(\borpeb) \Phi_{zz}(\Delta \borpe) = \beta \mu_{\xi},
\eeq
with $\nu_{n}$ denoting the $n-$dimensional analog of the thermal volume and $\Delta \borpe = \borpeb - \borpe$.  We have introduced an intralamellar  chemical potential $ \beta \mu_{\xi} = \beta \mu  - \sqrt{\ln \left ( \xi \nu_{1}/\pi) \right )} + 1/2$ that is  {\em smaller} than the corresponding value $\mu$ for the unlayered (isotropic or nematic) bulk fluid. 
Furthermore, $ \Phi_{zz}(\Delta \borpe)$ denotes a Mayer function pre-averaged over the out-of-plane fluctuations
\begin{align}
\Phi_{zz}(\Delta \borpe) &= \int_{-\infty}^{\infty} d z G(z) \int_{-\infty}^{\infty} d z^{\prime} G(z^{\prime}) \Phi(\Delta \bor) \nonumber \\
&= \int_{-\infty}^{\infty}  d \Delta z  {\mathcal G}(\Delta z)  \Phi(\Delta \bor),
\label{phizz}
\end{align}
where the distribution of the out-of-plane differential distance $\Delta z =  z^{\prime} -z $ is itself a Gaussian
\beq
{\mathcal G}({\Delta z})  = \left ( \frac{\xi}{2 \pi} \right ) ^{1/2} e^{-\frac{1}{2} \xi (\Delta z)^{2}}.
\label{ggauss}
\eeq
The key ingredient describing the rod interactions in the confined fluid is the second-virial coefficient for two rods with their $z-$coordinates constrained to a plane
\beq
\mathcal{K}_{0} = \int d\Delta \borpe \Phi_{zz}(\Delta \borpe).
\eeq
For purely hard rods, $\Phi = -1$ if the rod cores overlap and the kernel can be identified with (minus) the excluded volume between two rods with centres-of-mass confined to a bidimensional plane. This modification aside, the expression above is  identical to stationary condition obtained from  Onsager theory for conventional hard rod nematics \cite{onsager1949effects} with $\rho_{\perp}(\oma) = \rho_{\perp} f(\oma)$ in terms of some normalized orientational probability $f(\oma)$ that we will specify in Section IV. For later reference we also define Fourier transform (FT) of the excluded-volume manifold  between two  particles
\beq
\hat{\mathcal{K}}({\bf q}_{\perp}) = \int d\Delta \borpe \Phi_{zz}(\Delta \borpe ) e^{i {\bf q}_{\perp} \cdot \Delta \borpe},
\eeq
so that $\lim_{q_{\perp} \rightarrow 0} \hat{\mathcal{K}} = \mathcal{K}_{0}$. Fortunately, the double convolution of \eq{phizz} enables us to recast the kernel as a simple Fourier integral
\beq
\hat{\mathcal{K}}({{\bf q}_{\perp}})  = \int_{-\infty}^{\infty} \frac{d q_{z}}{2 \pi} [\hat{G}(q_{z})]^{2} \hat{\Phi}(q_{z} {\bf \hat{z}} + {\bf q}_{\perp}), 
\label{kernelft}
\eeq
with $\hat{G}(q_{z}) = \exp(-q_{z}^{2}/4 \xi) $ the FT of the Gaussian probability.
Realizing that the Mayer function simply designates (minus) the sub-volume where the cores overlap we write
\begin{align}
\hat{\mathcal{K}} ({{\bf q}_{\perp}} )  &= - \int d \Delta \bor_{\sigma} \int_{-\infty}^{\infty} \frac{d q_{z}}{2 \pi} e^{i {\bf q} \cdot \Delta \bor_{\sigma} - q_{z}^{2}/2\xi} \nonumber  \\
& =- \int d \Delta \bor_{\sigma} e^{i {\bf q}_{\perp} \cdot \Delta \bor_{\sigma}} {\mathcal G}(\Delta \bor_{\sigma} \cdot {\bf \hat{z}}). 
\label{kcore}
\end{align}
Next we invoke the standard particle-based parameterization of  this sub-volume via
$\Delta \bor_{\sigma} = \frac{L}{2}t_{1} \oma + \frac{L}{2} t_{2} \omb $ with coordinates $-1  < t_{i} < 1$ ($i=1,2$). The Jacobian associated with the coordinate transformation is given by $d \Delta \bor_{\sigma} = \frac{1}{2}L^{2}D | \sin \gamma | dt_{1} dt_{2}$ with $\gamma $ denoting the enclosed angle between the main rod axes.  This is our key result that we will elaborate below for two particular situations that are relevant for typical liquid crystal organizations of rod-shaped particles.

\section{Two limiting cases:  nematic and monolayer fluid}

Two limiting situations are easily identified from \eq{kcore}. First, if there is {\em no} lamellar confinement at all, the rod positions can explore full three-dimensional space such as in a nematic phase. Then  ${\mathcal G} =1$ and excluded volume to leading order for sufficiently slender rods $L/D \gg 1$ can be computed in analytical form   \cite{poniewierski1988nematic}
\beq
\hat{\mathcal{K}}^{\rm (nem)} = -2 L^{2}D | \sin \gamma |  j_{0}(\tfrac{L}{2} {\bf q} \cdot \oma )j_{0}(\tfrac{L}{2} {\bf q} \cdot \omb )  + \mathcal{O}(LD^{2}),
\eeq
in terms of the spherical Bessel function $j_{0}(x) = x^{-1} \sin x $. Since the focus of our study is not on nematics we do not discuss subsequent contributions of the order of the cylinder volume $LD^{2}$  (a technical discussion can be found in Ref. \cite{roij1996}). 

The second limiting case is more relevant to a smectic organization and occurs when the rods behave as a strict {\em monolayer} fluid with no positional fluctuations along the plane normal. The Gaussian distribution then becomes a delta distribution, i.e., $\lim_{\xi \rightarrow \infty} {\mathcal G}= \delta(\Delta \bor_{\sigma} \cdot {\bf \hat{z}})$. The hard-core kernel $\hat{\mathcal{K}}_{c}({\bf q}_{\perp})$ is then equivalent to the FT of the {\em excluded area} between two rods at equiplanar centres-of-mass ($\Delta z =0$). This quantity has been analyzed in detail in a study by Poniewierski \cite{poniewierski1988nematic}. Equating ${\mathcal G}= \delta(\Delta \bor_{\sigma} \cdot {\bf \hat{z}})$ in \eq{kcore}
we find
\begin{align}
\hat{\mathcal{K}}^{\rm (mono)} &= -\frac{L^{2}D}{2} |\sin \gamma |  \prod_{i=1}^{2} \int_{-1}^{1} dt_{i} \delta ( \Delta \bor_{\sigma} \cdot {\bf \hat{z}} ) e^{i {\bf q_{\perp} } \cdot \Delta \bor_{\sigma}} \nonumber \\
& = -L^{2}D | \sin \gamma | \frac{\sin \left (R/m \right )}{R } + \mathcal{O}(D^{2}),
 \label{kcfinal}
\end{align}
where $R$  is a lengthscale compiling the coupling between the wavevector and rod orientations
\beq
R=b_{z} ({\bf q}_{\perp} \cdot {\bf a}) - a_{z} ({\bf q}_{\perp} \cdot {\bf b}), 
\eeq
and ${\bf a} = L \oma/2$, ${\bf b} = L \omb/2$, $a_{z} = |{\bf a } \cdot {\bf \hat{z}}|$, $b_{z} = |{\bf b } \cdot {\bf \hat{z}}|$. The length $m$ selects the maximum of the latter two rod projections onto the lamellar normal $\hat{z}$ in terms of the Heaviside step function $\Theta$
\beq
m = a_{z}\Theta(a_{z} - b_{z}) +  b_{z}\Theta(b_{z} - a_{z}).
\eeq
The first contribution in \eq{kcfinal} is of order $\mathcal{O}(LD)$ and vanishes for a pair of perfectly parallel cylinders ($\gamma \rightarrow 0$). The next order term should be of $\mathcal{O}(D^{2})$  which we approximate here by the FT of a disc with radius $D$. We thus obtain a simplified expression for the total kernel for a monolayer rod fluid
\beq
 \hat{\mathcal{K}}^{\rm (mono)} \sim  -L^{2}D | \sin \gamma | \frac{\sin \left (R/m \right )}{R } - \pi D^{2} \frac{J_{1} (q_{\perp}D)}{\tfrac{1}{2}q_{\perp}D},
\label{kcorepc}
\eeq
with $J_{1}$ a Bessel function of the first kind. At present, there is no exact expression for the $\mathcal{O}(D^{2})$ term which should depend on the orientation of both rods and involve subtle correlations between the cylinder ends \cite{roij1996}.  We remark that the excluded volume for infinitely thin rods $(L/D \rightarrow \infty)$, given by the zero wavenumber limit of \eq{kcorepc}, is identical to the one derived by Oettel et al. \cite{oettel2016monolayers}. In real smectic phases, however, the rods are never perfectly confined to the lamellar mid-plane but are able to exercise small out-of-plane excursions that we will describe in Section V.

In the Appendix we demonstrate that our approach to describing excluded-volume interactions, namely pairing an orientationally-dependent contribution for infinitely thin particles with a tractable parallel core term, can be applied to both rods and discs alike and allows us to  accurately predict the onset of smectic or columnar freezing from a nematic reference fluid as well as the fluid-solid transition within a smectic monolayer.

\section{Orientational fluctuations}

We focus on the smectic A phase in which there is no long-ranged positional order transverse to the nematic director  and the one-body density can be written as $\rho_{\perp}(\oma) = \rho_{\perp} f(\oma)$ in terms of a fixed planar rod density $\rho_{\perp}$. For simplicity, we use the monolayer expression for the excluded volume \eq{kcfinal} and introduce the Helmholtz free energy of a lamellar fluid ignoring any term that is independent of the orientational probability $f$
\begin{align}
\frac{\beta F_{\rm L}}{N} \sim \langle \ln f(\oma) \rangle + \frac{8 \ell}{\pi} \phi_{\perp} g(\phi_{\perp}) \left \langle \left \langle   \frac{ |\sin \gamma |}{\bar{m} } \right \rangle \right \rangle.
\label{odf}
\end{align}
The brackets are short-hand for the angular integral $\langle \cdot \rangle = \int d \oma f(\oma) $  and likewise for a double angular average $\langle \langle \cdot \rangle \rangle = \int d \oma f(\oma) \int d \omb f(\omb) $. We further introduced the normalized lengths $\bar{m}(\oma , \omb) = m/(L/2)$,  planar rod packing fraction $\phi_{\perp} = (\pi/4)\rho_{\perp}D^{2} $ and rod aspect-ratio $\ell = L/D \gg 1$. 
Since the fluid residing within each lamella is spatially uniform, we apply a Lee-Parsons rescaling of the packing fraction $\phi_{\perp} \rightarrow \phi_{\perp} g(\phi_{\perp})$ where $g(\phi) = (1-\tfrac{3}{4} \phi)/(1- \phi)^{2}$ represents a correction factor to the second-virial approach, derived from the Carnahan-Starling equation of state for hard spherical particles,  that entails an \textit{ad-hoc} resummation of higher-order virial terms \cite{parsons1979nematic,lee1987numerical}. Despite the heuristic
nature of the theory, the Lee-Parsons formalism is known to give  quantitatively reliable results for the fluid phase behavior of elongated hard cylinders at elevated packing conditions \cite{mcgrother1996re}. We remark that the free energy above is very similar to that of a nematic phase except for the factor $\bar{m}$ which, as we will argue, will be very close to unity if the orientational fluctuations are weak. A crucial difference with the nematic fluid with prescribed density $\rho_{\rm nem}$, however, is that the planar packing fraction $\phi_{\perp}$ is not \textit{a priori} known but is controlled by the smectic spacing for a given  overall (3D) rod packing fraction as we will point out in Section VI. 

The formal way forward is to resolve the Euler-Lagrange equation from the extremum condition  $\delta F/ \delta f =0$ which can be done numerically \cite{herzfeld1984highly,van2005isotropic}.  In this study, we will pursue a much  simpler algebraic route to addressing intralamellar orientational order by assuming  orientational order to be both asymptotically strong and of simple uniaxial symmetry so that the angular probability only depends on the polar angle $\theta$ between the rod orientation vector $\oma$ and the layer normal along the $z$-axis.  Both criteria should be fulfilled for stable smectic structures composed of uniaxial rods whose interactions are apolar and achiral. Then, $f(\theta)$ can be approximated by a Gaussian trial function \cite{odijk1986theory}
\beq
f_{G}(\oma) \sim \frac{\alpha_{\perp}}{4 \pi} e^{-\frac{1}{2} \alpha_{\perp} \theta^{2}},
\label{fgauss}
\eeq
which applies within the interval $0 < \theta < \pi/2$ combined with its mirror form $f(\pi - \theta)$ for $\pi/2 < \theta < \pi$ and $\alpha_{\perp} \gg 1$ a variational parameter that is required to be large for consistency reasons given that the Gaussian trial form does not reduce to the required isotropic distribution  $f_{G} = (4 \pi)^{-1}$ in the limit $\alpha_{\perp} \rightarrow 0 $.   In a dense smectic environment, rod alignment is usually very strong and this criterion is readily met.  Ignoring constant terms we write the Helmholtz free energy as a combination of orientational and excluded-volume entropy per particle and keep only leading order terms  valid for small interrod angles. Then, $\bar{m} \rightarrow 1$ and the free energy \eq{odf} takes the following form
\begin{align}
\frac{F_{L}}{N} & \sim \langle \ln f_{G}(\theta) \rangle  + 4 \frac{ \ell }{\pi} \phi_{\perp} g(\phi_{\perp}) \langle \langle  \gamma  \rangle \rangle. 
\label{filgauss}
\end{align}
The double orientational averages are known up to leading order for large $\alpha_{\perp}$ \cite{odijk1986theory,vroege1992phase}
\begin{align}
\langle \ln f_{G}(\theta) \rangle &\sim \ln \alpha_{\perp} -1 \nonumber \\
\langle \langle  \gamma \rangle \rangle & \sim \sqrt{\pi / \alpha_{\perp}}. 
\label{gammav}
\end{align}
Minimizing the lamellar free energy $F_{\rm L}$ with respect to $\alpha_{\perp}$ then yields 
a quadratic relation with the planar rod packing fraction
\beq
\alpha_{\perp} \sim 4 ( \phi_{\perp} g(\phi_{\perp}) \ell)^{2}/\pi,
\label{alpha}
\eeq
similar to the case of a nematic fluid.

\section{Out-of-plane fluctuations}

We will now release the constraint of infinite lamellar confinement (monolayer case) and consider the density distribution $\rho({\bf r}) = \rho_{\perp} p(z)$ with intralamellar density $\rho_{\perp}$ and out-of-plane positional fluctuations along the layer normal described by a probability $p(z)$. Each rod is able to explore the full range of $z-$positions, subject to the constraint that the rod must not overlap with a neighboring particle whose mass center resides in the adjacent smectic layer a distance $\Delta_{S}$ away from the test rod. For simplicity, we assume the rods to be parallel and co-axial  so that overlap occurs whenever $| \Delta z | < L$.  We only consider nearest neighbor lamellar interactions. Then we may construct a fluctuation free energy per rod that depends on a single positional degree and reflects a balance between translational entropy and a volume-exclusion penalty
\begin{align}
& \frac{\beta \mathcal{F}_{\rm fluc}}{N}  = \int_{-\infty}^{\infty} dz p(z) [ \ln (\Lambda  p(z)) -1 ] + \frac{1}{2} \int_{-\infty}^{\infty} dz p(z) \nonumber \\ 
& \times \int_{-\infty}^{\infty} dz^{\prime} \left \{  p(z^{\prime} - \Delta_{S}) + p(z^{\prime} + \Delta_{S}) \right \} \Theta(L - |\Delta z|),
\label{freepos}
\end{align}
with $\Lambda$ the thermal wavelength. The corresponding single-particle distribution $p(z)$ can be resolved self-consistently through formal minimization $\delta \mathcal{F}_{\rm fluc}[p] /  \delta p =0$ which leads to
\begin{align}
 p(z) &= \exp \left ( \lambda^{\prime} -\int_{-\infty}^{\infty} dz^{\prime}  p(z^{\prime} ) \right . \nonumber \\ 
 & \times \left . \{ H(L - |\Delta z  - \Delta_{S}|) + H(L - |\Delta z + \Delta_{S}|) \} \right ),
 \label{pzsc}
\end{align}
where $\lambda^{\prime}$ denotes a multiplier ensuring normalization $\int dz p(z) =1 $. If we assume the rods in the neighboring lamellae to be fixed at their mid-plane, we can establish an analytical mean-field solution  from $p(z^{\prime}) \approx \delta(0)$ which gives $p(z) = 1/2(\Delta_{S} -L)$  for $|z| < (\Delta_{S} - L)$ and $p(z)=0$ otherwise. A numerical solution for $p(z)$ can be obtained through iteration but the results merely lead to entropic rounding at the edges of the step function thus rendering the distribution continuously differentiable.  Even though $p(z)$ is not Gaussian,  we may tentatively identify $p(z) = G(z)$ via \eq{geezer} and connect the mean-squared out-of-plane displacement to the Gaussian parameter $\xi$
\beq
\xi \sim \frac{3}{2(\Delta_{S} -L)^{2}}, 
\label{xidus}
\eeq
revealing, as expected, that out-of-plane fluctuations $\sim \xi^{-1}$ grow stronger at larger lamellar distances $\Delta_{S} >L$.  

For real smectic phases such as those composed of {\em fd} rods,  the out-of-plane fluctuations and the typical energy barrier the rods need to overcome to jump from one layer to the next have been studied in detail \cite{lettinga2007self, pouget2011dynamics,Alvarez2017,Chiappini2020}. The  density distribution normal to the layer for small deviations from the lamellar  mid-plane can be described by a Gaussian
\beq
p(z) \approx \exp \left [ -  2 \pi^{2} U_{0} (z/L)^{2} \right ] \hspace{0.2cm} z/L \ll 1, 
\eeq
with $U_{0}$ a typical smectic layer potential (expressed in units $k_{B}T$) for which experimental values are available (see Section IX). It should be noted that the non-Gaussian tail of $p(z)$ plays an essential role in determining the potential amplitude. 
Comparing with our Gaussian distribution we may identify the smectic potential to the layer spacing for small out-of-plane excursions $z/L \ll 1$
\beq
U_{0} \sim \frac{3}{4 \pi^{2}} \left (\frac{\Delta_{S}}{L} -1 \right )^{-2}.
\eeq
In Section X we will discuss a comparison with experimental measurements of the smectic potential for {\em fd} rods.

\subsection{Effect of layer interdigitation}

The naive description proposed so far overestimates the smectic potential since rods are known to migrate from one layer to the next  \cite{lettinga2007self, pouget2011dynamics} which suggests a degree of {\em dynamic interdigitation} between smectic layers. This can be understood from the fact that rods are never closely packed together within each lamella but experience a certain degree of free volume that enables particles to at least partially penetrate into a neighbour layer. Clearly, interdigitation becomes less prominent at larger packing conditions.  This effect can be taken into account, at a simplified mean-field level, by introducing  a finite layer interdigitation  potential $W$ which broadens the out-of-plane distribution $p(z)$ and partly captures the non-Gaussian tails of the out-of-plane distribution which now reads
\beq
 p (z) =
\begin{cases}
1 &  |z| < \Delta_{S} - L \\
e^{-\beta W }  & \Delta_{S} - L  < | z| < \Delta_{S} - L + \zeta \\
0 & | z| > \Delta_{S} - L + \zeta, 
\end{cases}
\label{pzrod}
\eeq
with $0< \zeta < L$ a interdigitation depth such that  $\zeta =L$ implies that each rod can fully penetrate the adjacent layer while for $\zeta =0 $ no interdigitation occurs at all and the previous result \eq{pzsc} is recovered. In real systems with strong in-plane crowding, we expect $\zeta$ to be only a fraction of the rod length.  After some algebra, we find that the smectic layer potential associated with $p(z)$ reads
\beq
U_{0} \sim \frac{3[(\bar{\Delta}_{S} -1) + e^{-\beta W}\bar{\zeta}] }{4 \pi^{2} [(\bar{\Delta}_{S} -1)^{3} + e^{-\beta W}((\bar{\Delta}_{S} -1 +\bar{\zeta})^{3}-(\bar{\Delta}_{S} -1)^{3}) ]},
\eeq
in terms of the interdigitation depth $\bar{\zeta} = \zeta /L$   and spacing $\bar{\Delta}_{S} = \Delta_{S}/L$ both  normalized to the rod length. The insertion potential can be interpreted as a potential of mean force represented by the free energy cost of inserting a rod into a bidimensional fluid for which Scaled Particle Theory (SPT) can be invoked \cite{reiss1959statistical,hansen1988theory}. Then, $W$ is simply the excess chemical potential $\mu_{\rm ex}$ of a bidimensional fluid of hard discs representing the rod cross sections residing on the lamellar area $A$. This expression is well known and reads \cite{helfand1961theory}
\beq
\beta W = - \ln ( 1- \phi_{\perp}) + \frac{1 + \phi_{\perp}(1- \phi_{\perp}) }{(1- \phi_{\perp})^{2}}, 
\eeq
where $\phi_{\perp}$ denotes the in-plane packing fraction that implicitly depends on the lamellar spacing via $\phi_{\perp} = \phi \Delta_{S}$. In fact, the SPT theory and variations thereof \cite{baus1987thermodynamics} have been used routinely in cell-type theories for smectic and other liquid crystal states \cite{hentschke1989equation,graf1997cell,graf1999phase,wensink2004equation,peters2020algebraic}.
The presence of layer interdigitation lowers the smectic potential to values that are in good agreement with experimental measurements on smectic phases of {\em fd} rods as we will discuss in Section X.

\section{From monolayer fluid to smectic liquid crystal}

 We now wish to construct a tractable thermodynamic theory for a 3D smectic phase composed of $M \rightarrow \infty$ lamellae. For purely hard interactions an appealing route is to use cell-theory to correlate the 2D fluid behavior with a 1D periodicity of the smectic liquid  crystal.

 The basic assumption of the cell approach for describing a smectic structure is that the vertical degrees of each rod (along the layer normal) are constrained by the presence of the adjacent smectic layers such that rods are not permitted to cross layer boundaries. Maximum positional freedom of each rod is then dictated by the size of the cell which is proportional to the smectic layer spacing. With these constraints, the 1D ordered dimension and the 2D fluid dimensions of the smectic structure are fully decoupled and can be modelled separately.  Similar arguments can be applied to other positionally ordered liquid crystals, such as the columnar, and even solids \cite{hentschke1989equation}.

 Focusing on the 1D ordered dimension, we assume that the layers impart some effective potential that can be purely hard-core like \cite{hentschke1989equation,graf1997cell} or feature an additional continuous potential to account for long-range repulsive interactions between particles \cite{han1996avoidance, kramer1999avoidance}. Then, the excess free energy is connected to the mean out-of-plane free distance $\Delta_{S}$ each rod is able to explore before touching a rod from an adjacent layer. The configurational integral of a collection of $M$ uncorrelated cells reads \cite{barker1976liquid}
\beq
Q \approx \left( \Lambda^{-1} \int d z e^{-\beta u_{\rm cell}} \right )^{M},
\eeq
where the cell potential can be reverse-engineered from the out-plane-fluctuation probability \eq{pzrod} through a simple Boltzmann inversion $u_{\rm cell}(z) = -\beta^{-1} \ln p(z)$
\beq
u_{\rm cell} (z) =
\begin{cases}
0 &  |z| < \Delta_{S} - L \\
 W   & \Delta_{S} - L  < | z| < \Delta_{S} - L + \zeta \\
\infty & | z| > \Delta_{S} - L + \zeta.
\end{cases}
\label{ucell}
\eeq
The soft potential $W$ accounts, albeit heuristically, for the degree of rod interdigitation across adjacent lamellae. 
The excess free energy per rod $\beta F_{\rm cell}=- \ln Q$ associated with smectic layering in the thermodynamic limit $M \rightarrow \infty$ thus reads   
\beq
\frac{F_{\rm  cell}}{N} \sim -\ln [\Lambda^{-1} ( \Delta_{S} - L + \zeta e^{-\beta W})] - \ln2.
\eeq
The contribution proportional to the interdigitation depth $\zeta$ is due to the {\em soft-cell} potential \eq{ucell} and is not considered in conventional cell approaches where  this effect is ignored, i.e.,  $W \rightarrow \infty $ ($\zeta=0$).  Conservation of number of particles and the single-occupancy condition (each rod belongs to only one smectic layer)  requires that the overall (3D) rod density  relate to the in-plane concentration via $\rho = \rho_{\perp}/\Delta_{S}$.

\begin{figure}
	\includegraphics[width = \columnwidth]{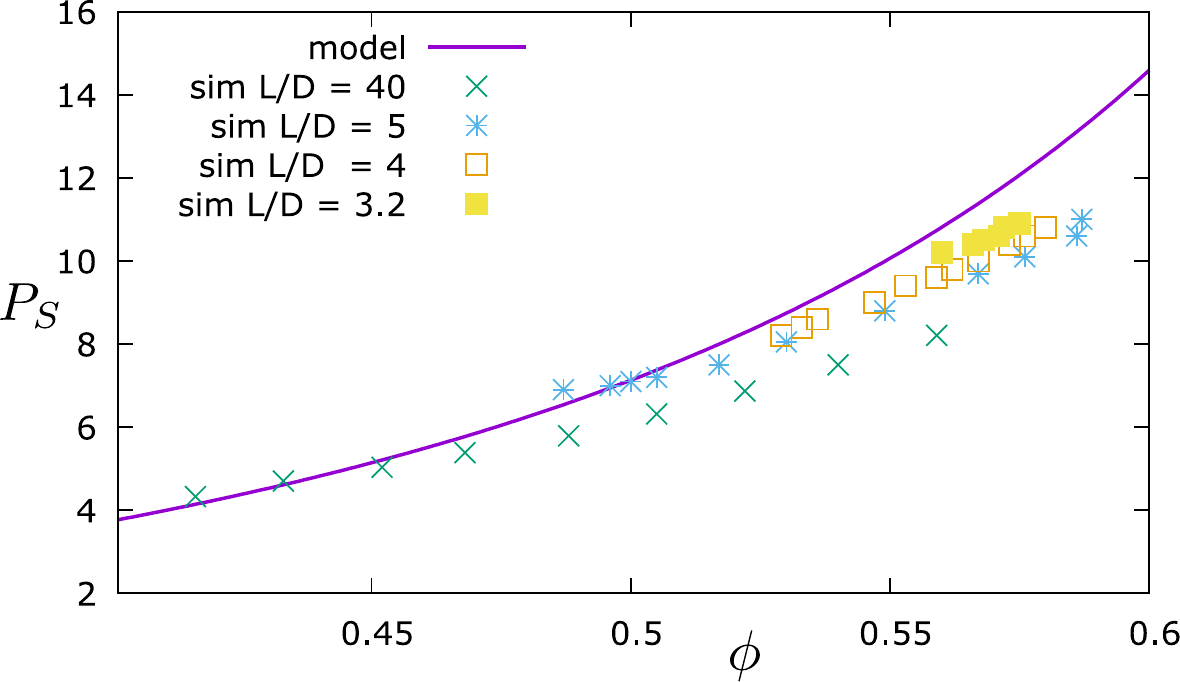}
	\caption{Equation-of-state of a Smectic A phase 
 of hard rods. Plotted is the osmotic pressure $P_{S}$ (in units $k_{B}T$ per rod volume) versus the packing fraction $\phi$.  The prediction from the theoretical model is generic and does not depend on the rod aspect ratio $\ell=L/D$. Simulation results are quoted from McGrother \textit{et al.} \cite{mcgrother1996re} ($L/D = 3.2,4,5$) and Bolhuis \textit{et al.} \cite{bolhuis1997tracing} ($L/D=40$).     }
	\label{seos}
\end{figure}

 We now wish to put our model to a quantitative test by comparing the equation-of-state of the (3D) smectic phase for hard cylinders to results from computer simulation for hard spherocylinders. Subtle end-cap effects that distinguish these two particle shapes are deemed of minor importance for sufficiently elongated particles $L/D \gg 1$.   
In order  to address the thermodynamics of the smectic phase we combine the above cell contribution with the intralamellar fluid contribution discussed previously [cf. \eq{odf}]. Ignoring all constant factors we find a simple expression that incorporates the ideal, orientational, rod excluded-volume and lamellar entropies, respectively
\begin{align}
\frac{F_{\rm S}}{N} & \sim \ln \phi  + 2 \ln [ \phi_{\perp} g(\phi_{\perp})]+ 2 \phi_{\perp} g(\phi_{\perp})\nonumber \\ 
& - \ln \left [ 1 - \bar{\Delta}_{S}^{-1}( 1-  \bar{\zeta}  e^{-\beta W}) \right ],
\label{fsma}
\end{align}
where $\bar{\Delta}_{S} = 2 \pi/q_{S}L$ denotes the spacing (in units rod length)  between  adjacent smectic layers. In the Gaussian approximation, the orientation-dependent part of the excluded volume produces a constant of ``2'' as originally pointed out by Odijk \cite{odijk1986osmotic}.   The equilibrium spacing is then easily found from minimization $\partial F_{S}/ \partial \Delta_{S} =0$ and the pressure from $P_{S} = \phi^{2}[ \partial (F_{S}/N)/\partial \phi] $.  The results  in \fig{seos} demonstrate that our simple model gives a quantitatively reliable prediction of the equation-of-state at relatively low packing fraction ($\phi < 0.5$) while somewhat overestimating the simulated values at large packing fraction. We find that the pressure is rather insensitive to the choice of the interdigitation depth $\zeta$. This confirms the robustness of the cell description 
in the sense that the thermodynamic properties of the smectic phase are  insensitive to the details of the cell potential. However, we will demonstrate in the next Section that the mechanical properties of the smectic material {\em do} depend on  the out-of-plane fluctuations which in turn are steered by the degree of layer interdigitation and the choice of $u_{\rm cell}$.

\section{Layer compression modulus}

\begin{figure}
	\includegraphics[width = \columnwidth]{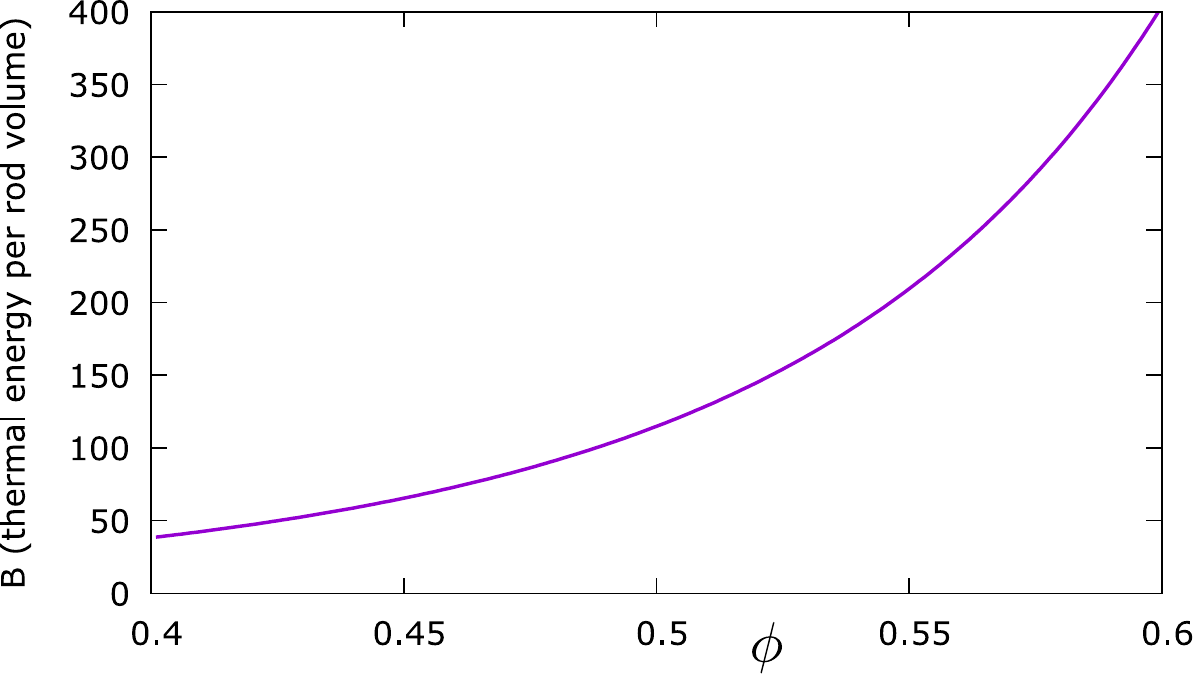}
	\caption{ Layer compressibility modulus $B$ of a hard-rod smectic as a function of the overall packing fraction $\phi$. The modulus is expressed in units $k_{B}T/v_{r}$ in terms of the rod volume $v_{r} = \tfrac{\pi}{4} LD^{2}$.   }
	\label{comp}
\end{figure}

In order to describe long-wavelength elasic distortions of a colloid-based smectic material we consider 
the elastic free energy per unit volume of a smectic liquid crystal which is given by the following expression \cite{de1993physics}
\beq
\frac{F_{\rm S}}{V} = \frac{B}{2} \left [  \frac{\partial u }{\partial z}    - \frac{1}{2} \left (\frac{\partial u}{\partial x} \right )^{2} \right ]^{2} + \frac{K_{1}}{2} \left ( \frac{\partial ^{2} u}{\partial x^{2}} \right )^{2},
\eeq
with $u(x,z)$ the displacement field denoting a local displacement of the smectic layer away from its equilibrium position. The deformations are described by two elastic moduli;  the layer compression modulus $B$ and splay modulus $K_{1}$ which is associated with layer bending which in the above expression is assumed to be uni-directional along the $x-$axis (see \fig{sbend}). Both moduli can be  combined into a single length scale; the smectic penetration length $\lambda = \sqrt{K_{1}/B}$. Furthermore, the Landau-Peierls instability dictates that the mean-squared fluctuations of the displacement field $u$ diverge with the system size $\ell_{s}$ \cite{de1993physics,dejeu2003structure} 
\beq
\langle u({\bf r})^{2} \rangle  = \frac{k_{B}T}{8 \pi \sqrt{K_{1}B}} \ln (\ell_{s} /\Delta_{S}),
\eeq
with $\sqrt{K_{1}B}$ a characteristic energy scale per unit  area which along $\lambda$ defines the elastic response of a smectic material. We now wish to quantify these for the specific case of hard rod smectic by analyzing each of the contributions separately.   First, the compression modulus $B$ can be defined following a general analysis of Ref. \cite{ligoure1997smectic} for bilayer smectics with spacing $\delta_{S}$
\beq
B = \delta_{S} \left ( \frac{\partial^{2} f_{L}}{\partial \bar{\delta}_{S}^{2}} \right )_{\mu},
\eeq
with $\bar{\delta} = \delta_{S} - h$ and $h$ the bilayer thickness. The free energy $f_{L}$ per bilayer and unit area must be defined at constant chemical potential of the bilayer constituents in order to facilitate fluctuations in the surface coverage upon dilation of compression of the smectic layers.  Translating this expression to our current entropy-stabilized hard rod smectic with spacing over rod length $\bar{\Delta}_{S}$ we find that the layer compressibility modulus (rendered dimensionless in units of thermal energy $k_{B}T$ per particle volume $v_{r}$) reads
 \beq
B = \bar{\Delta}_{S} \left ( \frac{\partial^{2} [\phi_{\perp} F_{S}/N ]}{\partial \bar{\Delta}_{S}^{2}} \right )_{\phi}.
\eeq
The difference with the previous definition for the bilayer case is that layer compression happens at {\em constant} overall rod packing fraction  instead of chemical potential and that the average layer thickness is fixed at $L$. This seems a reasonable assumption for most smectic organizations, including bilayer phases. Our prediction of the compressibility modulus is shown in \fig{comp} and features a fairly steep (about four-fold) increase of the compression resistance of a hard-rod smectic at moderately high   densities $0.4 < \phi < 0.55$.

\section{Splay modulus}

The  modulus $K_{1}$ refers to a splay deformation of the layer normal, which coincides with the orientational director of the  rods, upon layer bending. To the best of our knowledge, no attempt has been made so far to quantify $K_{1}$ from microscopic theory.  For {\em nematic} phases, however, theoretical predictions of the Frank elastic moduli have been proposed decades ago starting with the work of Straley \cite{straley1976theory}. Within the second-virial approximation the splay modulus is formally given by the following expression \cite{straley1976theory,allen1993hard}
\beq
K_{1}  = k_{B}T \frac{\rho^{2}}{2} g(\phi) \left \langle \left \langle  \int d {\bf \Delta r}  (\Delta x)^{2}  \Phi( {\bf \Delta r} ) \omega_{x} \omega^{\prime} _{x} \right \rangle \right \rangle_{\dot{f}},
\eeq
with $\dot{f}$ denoting the derivative of the orientation distribution with respect to its argument. Using an asymptotic expansion  based on the trial form \eq{fgauss} Odijk \cite{odijk1986elastic} obtained the follow scaling expression, valid in the limit of strong nematic alignment
\beq
\beta K_{1} D \sim \frac{7 }{8 \pi} \phi \ell,
\label{K1nem}
\eeq
with $\ell=L/D$ the rod aspect ratio. 
In the smectic phase, the rod positions are not uniformly distributed but strongly localized about the lamellar mid-plane. 
The corresponding microscopic expression for the splay modulus of a smectic phase reads
\begin{align}
\beta K_{1}  & \sim \frac{\rho^{2}}{2} g(\phi_{\perp}) \Delta_{S} \left \langle \left \langle  \int d {\bf \Delta r_{\perp}}  (\Delta x)^{2}  \Phi_{zz} \omega_{x} \omega^{\prime}_{x} \right \rangle \right \rangle_{\dot{f}} \nonumber \\
& \sim   -\frac{\rho^{2}}{2} g(\phi_{\perp}) \Delta_{S}  \left \langle \left \langle  \int d {\bf \Delta r_{\sigma}} \mathcal{G}(\Delta  z_{\sigma}) (\Delta  x_{\sigma})^{2}  \omega_{x} \omega^{\prime} _{x} \right \rangle \right \rangle_{\dot{f}}.
\end{align}
Analogous to the  case of a nematic phase we may obtain a scaling expression of the double orientational average using the Gaussian algebraic manipulations outlined in  Ref. \cite{odijk1986elastic}. The key step is to eliminate the azimuthal part $\cos 2\Delta  \varphi = 2 \cos^{2} \Delta \varphi -1$ via $\gamma^{2} \sim \theta^{2} + \theta^{\prime 2} - 2 \theta \theta^{\prime} \cos \Delta \varphi$ and apply double Gaussian averages over the remaining combinations of $\gamma $ and $\theta_{1}$ and $\theta_{2}$ that are listed in \cite{odijk1986elastic}. A tedious but straightforward computation then leads to following result up to leading order for weak out-of-plane fluctuations ($\xi L^{2} \gg 1$)
\beq
\beta K_{1} D \sim \frac{11 }{2 \pi^{3/2}} \alpha_{\perp}^{-1/2} (\phi \ell)^{2} g(\phi_{\perp}) \left ( 1 - \frac{6}{ \sqrt{2 \pi \xi L^{2}}}  \right ).
\label{splayalpha}
\eeq
Higher order corrections are of $ \mathcal{O}(\xi^{-1})$ and will not be considered here. Plugging in the quadratic relationship for $\alpha_{\perp}$ [\eq{alpha}]  we obtain the following analytical result 
\beq
\beta K_{1} D \sim \frac{11 }{4 \pi} \left ( 1 - \frac{6}{ \sqrt{2 \pi \xi L^{2}}}  \right ) \phi \ell.
\label{psplay}
\eeq
We infer from comparison with Eq.~\ref{K1nem} that in the absence of positional fluctuations ($\xi \rightarrow \infty$) the splay elasticity is about three times larger than that of a nematic phase at comparable rod concentration $\phi \ell $.  As expected, the presence of out-of-plane fluctuations  with strength $\xi^{-1}$  reduces the splay modulus and will only reach the nematic level for very weakly ordered smectics with a large spacing and strong positional fluctuation about the lamellar mid-plane. For a typical hard rod smectic  with a spacing of about $\Delta_{S} = 1.1L$, equivalent to $\xi L^{2} \sim \mathcal{O}(10^{2})$,  $K_{1}$ is about 15 \% smaller than in the limit of infinite lamellar confinement ($\xi \rightarrow \infty$).

\begin{figure}
	\includegraphics[width = \columnwidth]{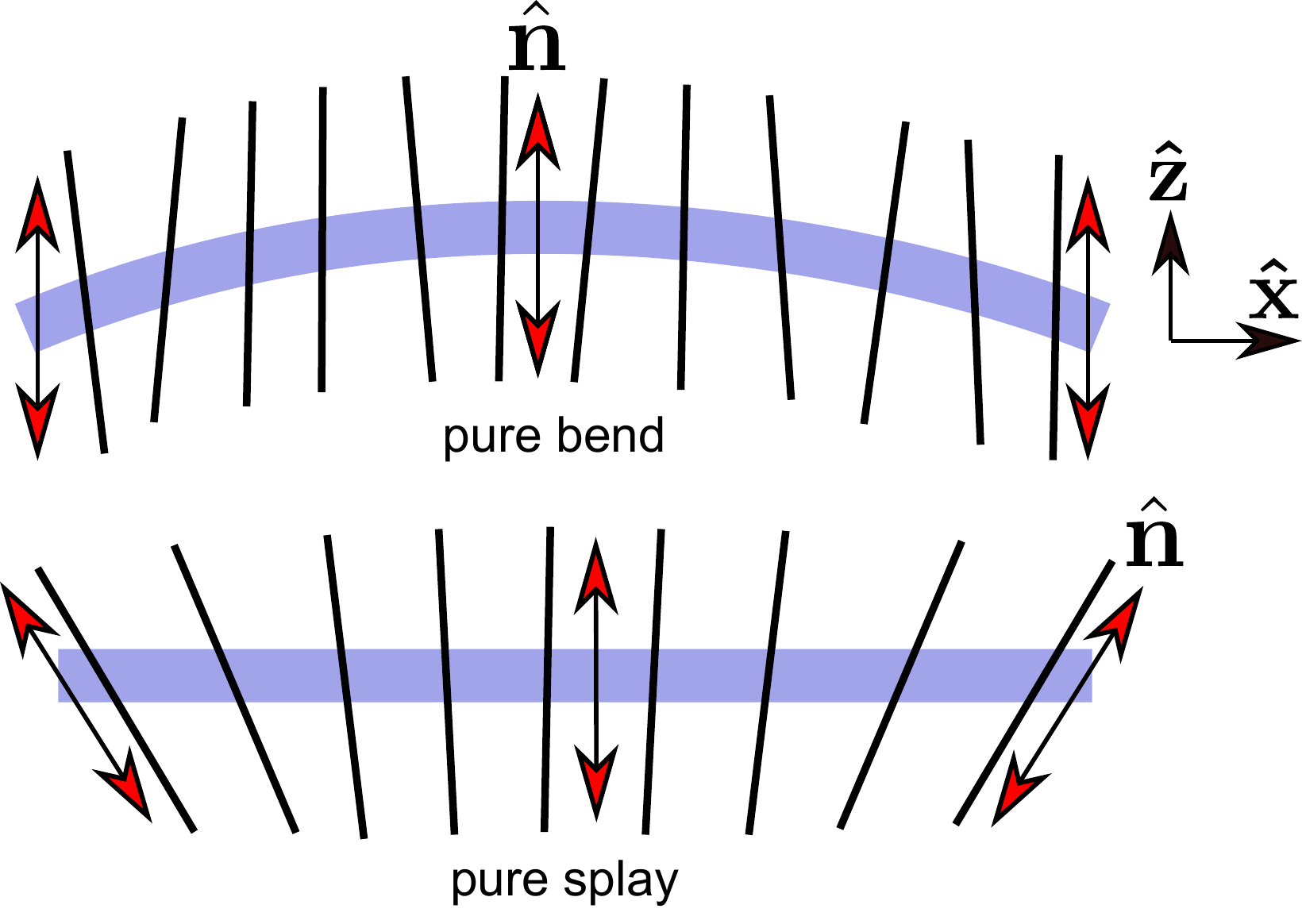}
	\caption{Two principal effects of layer bending; (top) a pure bend deformation of the smectic layer keeping a uniform (splayless) director field  and (bottom) a splay deformation of the director field without layer bending.  The lamellar mid-plane is indicated in blue.}
	\label{sbend}
\end{figure}

So far we have only considered pure splay and ignored any bending of the layers. 
Let us now contemplate an infinitesimal bending of the smectic layer while keeping a {\em uniform} director field ${\bf \hat{n}}({\bf r}) = {\bf \hat{z}}$. This amounts to bending a smectic layer without splaying the director field as illustrated in \fig{sbend}.  The excluded volume \eq{kcfinal} between two test rods with centres-of-mass confined on a weakly curved 2D plane now depends on the  curvature $\kappa$ which should be weak on the scale of the particle size ($\kappa L \ll 1$).
 The projected excluded volume  (with dimension area) between two thin hard rods with centres-of-mass residing on a curved (2D) layer is given by minus the kernel $\mathcal{K}$ and reads
\begin{align}
\mathcal{ K}(\kappa) & \sim -\tfrac{1}{2} L^{2}D |\sin \gamma |  \prod_{i=1}^{2} \int_{-1}^{1} dt_{i} \nonumber \\
 & \times \mathcal{G} [ \Delta z_{\sigma}  - \kappa^{-1}(\sqrt{1- (\kappa \Delta x_{\sigma})^{2}} -1)]. 
 \label{kcq}
\end{align}
Taylor expanding up to second order in the curvature $\kappa$ formally gives
\beq
\mathcal{K}(\kappa) \approx \mathcal{K}(0) + \kappa^{2} \delta \mathcal{K} + \mathcal{O}(\kappa^{4}).
\eeq
We are primarily interested in the  correction  $\delta \mathcal{K}$ which has dimension length to the fourth power. The double integral \eq{kcq} can be rendered analytically tractable by keeping the lowest order term for small angular fluctuations $\theta \ll 1 $ and positional fluctuations about the smectic plane $\xi L^{2} \gg 1$ we find
\beq
\delta \mathcal{K} \sim  \frac{3}{128 \sqrt{2 \pi}} L^{5} D  \xi  | \gamma |^{5}.
\eeq
The elastic free energy density  corresponds to the change in free energy per unit volume incurred by  slight layer bending.  Within the second-virial approximation it can be written as
\beq
\frac{ \delta F_{\rm S}}{V}  \approx \tfrac{1}{2}k_{B}T \rho^{2} g(\phi_{\perp}) \Delta_{S}  \langle \langle - \delta \mathcal{K}  \rangle \rangle \kappa^{2}.   
\eeq
The brackets denote a double  average over the  orientational distribution function which we assume is unaltered by layer bending. We may read off the bend modulus under the constraint of a uniform
director field
\beq
K_{b}  \sim k_{B}T \rho^{2} g(\phi_{\perp}) \Delta_{S} \langle \langle - \delta \mathcal{K} \rangle \rangle.
\eeq
It is easily ascertained that this quantity has units of force.       
Using the Gaussian orientational average $ \langle \langle  | \gamma |^{5} \rangle \rangle  \sim 60 \pi^{1/2} \alpha_{\perp}^{-5/2}$ \cite{odijk1986elastic} up to leading order for strong in-plane alignment $\alpha_{\perp} \gg 1$ combined with \eq{alpha} 
we  find the following scaling expression for the bend modulus
\begin{align}
\beta K_{b}  D & \propto  -(\xi L^{2}) \alpha_{\perp}^{-5/2} \nonumber \\
& \propto  - (\phi \ell)^{-3} g(\phi_{\perp})^{-4}\Delta_{S}^{-4} (\xi L^{2}).
\end{align}
We observe that the layer stiffness is much more sensitive to the strength of the orientational fluctuations $\alpha_{\perp}$ than to the out-of-plane fluctuations which are principally controlled by the lamellar distance $\Delta_{S}$ via \eq{xidus}.
Noting  that $\phi \ell \gg 1$ for thermodynamically stable smectics and $\xi L^{2} \sim \mathcal{O}(10^{2})$ for typical lamellar spacings we find that for moderate rod aspect ratio $\ell =10$ the  bend contribution is at  least two orders of magnitude smaller much than the splay modulus \eq{psplay}. The difference is even greater for slender rods $\ell \gg 10$ where splayless layer bending  has a marginal effect on the smecto-elastic response. 
    We remark that $K_{1}$ depends non-trivially on the packing fraction  because the equilibrium spacing $\Delta_{S}$ itself depends non-analytically on $\phi$ though the minimum condition $\partial F_{\rm S}/\partial \bar{\Delta}_{S} = 0$ of the smectic free energy \eq{fsma}.

\begin{figure}
	\includegraphics[width = \columnwidth]{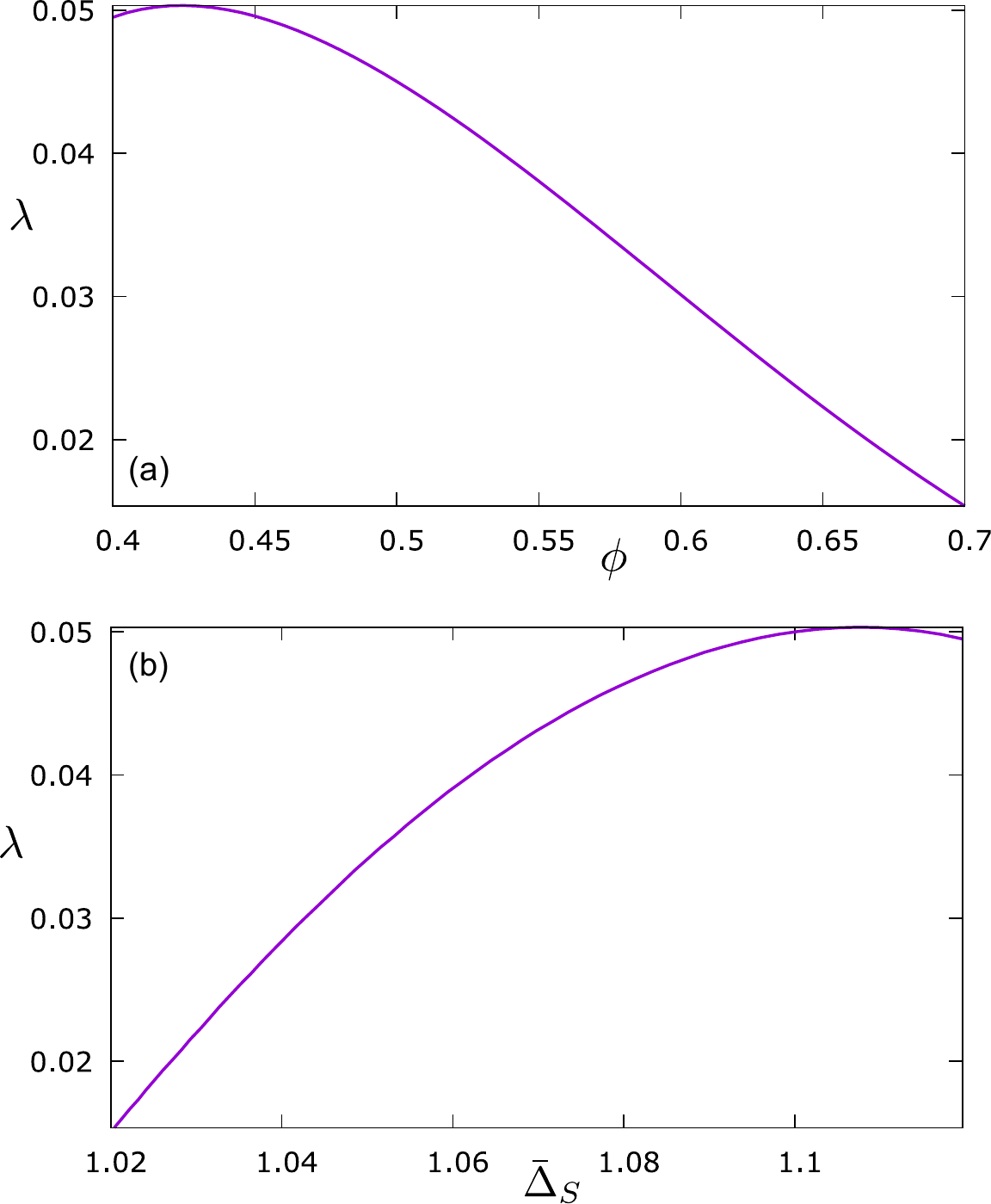}
	\caption{ (a) Penetration length $\lambda = \sqrt{K_{1}/B}$ of a hard rod smectic  in units of the rod length $L$ versus the rod packing fraction $\phi$. The penetration length is shape-independent and does not depend on the rod aspect ratio. (b) Same as a function of the smectic layer spacing $\bar{\Delta}_{S}$ (in units rod length $L$).  }
	\label{pene}
\end{figure}

  Now that we have established that director splay is much more important than pure layer bending, at least for hard-rod smectics,  we may combine the prediction for the compressibility modulus $B$ with the expression for the intralamellar splay elastic constant \eq{psplay} and compute the smectic penetration length $ \lambda = \sqrt{ K_{1}/B}$  shown in \fig{pene}. We find that this length constitutes only a tiny fraction of the rod length (or smectic layer spacing) and that it decreases linearly at elevated packing conditions.  We further observe that the penetration length exhibits a (weak) maximum around $\phi \approx 0.43$. We have not verified the robustness of the maximum in relation to the  approximations made in our theory.

 \section{Role of backbone flexibility}

In order to render our theory more appropriate for experimental systems such as filamentous virus {\em fd} rods we have to account for the slight flexibility of the virus backbone, which has been shown to quantitatively affect the phase diagram \cite{Barry2009,grelet2014hard}. Liquid crystals of semi-flexible polymers have also been the subject of recent simulations in an effort to test  scaling concepts that were put forward in earlier theoretical work on polymeric liquid crystals (see Refs. \cite{milchev2018nematic,milchev2019smectic} for a discussion).

In our theory we shall account for backbone flexibility  by introducing a correction to the orientational entropy of perfectly rigid rods considered thus far in our modelling.  The presence of rod flexibility, however slight, gives rise to an additional entropy generated by the internal configurations of a so-called worm-like chain (wlc) \cite{vroege1992phase} which can be described by the following non-linear expression in the orientational probability $f$  \cite{khokhlov1982liquid,kuriabova2010linear}
\beq
  \frac{F_{wlc}}{N} = - \frac{2L}{3\ell_{p}}  \int d \oma [ f( \oma)]^{1/2} \nabla^{2}  [ f(\oma)]^{1/2}.
  \label{wlc}
\eeq
Here, $\nabla^{2} $ denotes the Laplace operator on the unit sphere and $\ell_{p}$ is the persistence length that represents the  typical length scale over which the orientational fluctuations of the local (Kuhn) segments of each rod are correlated.  Since  {\em fd} rods are rather stiff with a persistence length strongly exceeding the contour length ($\ell_{p} \sim 3 -10L $) \cite{Barry2009}, the entropy associated with the internal fluctuations of the effective segments is relatively small compared to the orientational entropy of the entire rod. In the semi-flexible regime ($D \ll \ell_{p} \ll L$) it can be demonstrated that the second-virial coefficient between two worm-like chains between is the same as the one for rigid rods \cite{odijk1986theory, vroege1992phase}. Strictly, this is no longer the case in the rod limit $\ell_{p} \gg L$ where rod-rod exclusion is expected to depend on the degree of flexibility but we shall assume deviations from the rigid rod excluded volume to be small for the case of {\em fd}.    

The worm-like chain entropy can be estimated from Gaussian orientational distribution $f_{G}$ and reads up to leading order for strong orientational order \cite{vroege1992phase}
\begin{align}
\int d \oma  [ f_{G}( \oma)]^{1/2} \nabla^{2}  [ f_{G}(\oma)]^{1/2} & \sim -\frac{\alpha_{\perp}}{2}. 
\end{align}
From which we obtain
\beq
  \frac{F_{wlc}}{N} \sim  \frac{L \alpha_{\perp}}{3\ell_{p}}.  
\eeq
 In case of strong in-plane orientation order ($\alpha_{\perp} \gg 1$) the number of polymer conformations is severely limited which thus leads to a free energy penalty.  Adding the worm-like chain correction to the Helmholtz free energy per particle \eq{filgauss} for the lamellar fluid we find (ignoring irrelevant constants) 
\begin{align}
\frac{F_{\rm L}}{N} & \sim  \ln \alpha_{\perp}   +  4 \frac{ \ell }{\pi} \phi_{\perp} g(\phi_{\perp}) \sqrt{\pi / \alpha_{\perp}} + \frac{L \alpha_{\perp}}{3\ell_{p}},  
\label{fgaussflexi}
\end{align}
which upon minimization yields a cubic equation in $\alpha_{\perp}$
\beq
\alpha_{\perp}^{1/2}  \sim (2/\pi^{1/2}) \phi_{\perp} g(\phi_{\perp}) \ell - \frac{L}{3 \ell_{p}} \alpha_{\perp}^{3/2}.  
\eeq
The physical branch is the one for which $\alpha_{\perp}$ is real-positive and which converges to  \eq{alpha}  in the rigid rod limit $\ell_{p}/L \rightarrow \infty$. Compared to the case of perfectly rigid rods, a small but finite amount of backbone flexibility leads to a reduction of the  orientational order of the rods for any given planar packing fraction $\phi_{\perp}$ and rod aspect ratio $\ell$.  

We may now simply repeat the analysis we did for the rigid rods by retaining the intralamellar orientational order parameter $\alpha_{\perp}$ to be resolved from the cubic equation, shown previously. 
The total free energy of the smectic phase, again ignoring irrelevant constants, is now given by
\begin{align}
\frac{F_{\rm S}}{N}  \sim & \ln \phi  +  \ln \alpha_{\perp} + \phi_{\perp} g(\phi_{\perp}) \left ( \frac{4  \ell }{(\pi \alpha_{\perp})^{1/2}} + 2 \right ) \nonumber \\  
 & +  \frac{L \alpha_{\perp}}{3 \ell_{p}}  -
 \ln \left [ 1 - \bar{\Delta}_{S}^{-1}( 1-  \bar{\zeta}  e^{-\beta W}) \right ].
\label{fsmaflex}
\end{align}
From the equation above, we identify the entropies associated with ideal gas behavior, orientational fluctuations, excluded-volume repulsion (composed of a parallel core plus an orientation-dependent part), single-rod conformational fluctuations and lamellar confinement, respectively.

\begin{figure}
	\includegraphics[width = \columnwidth]{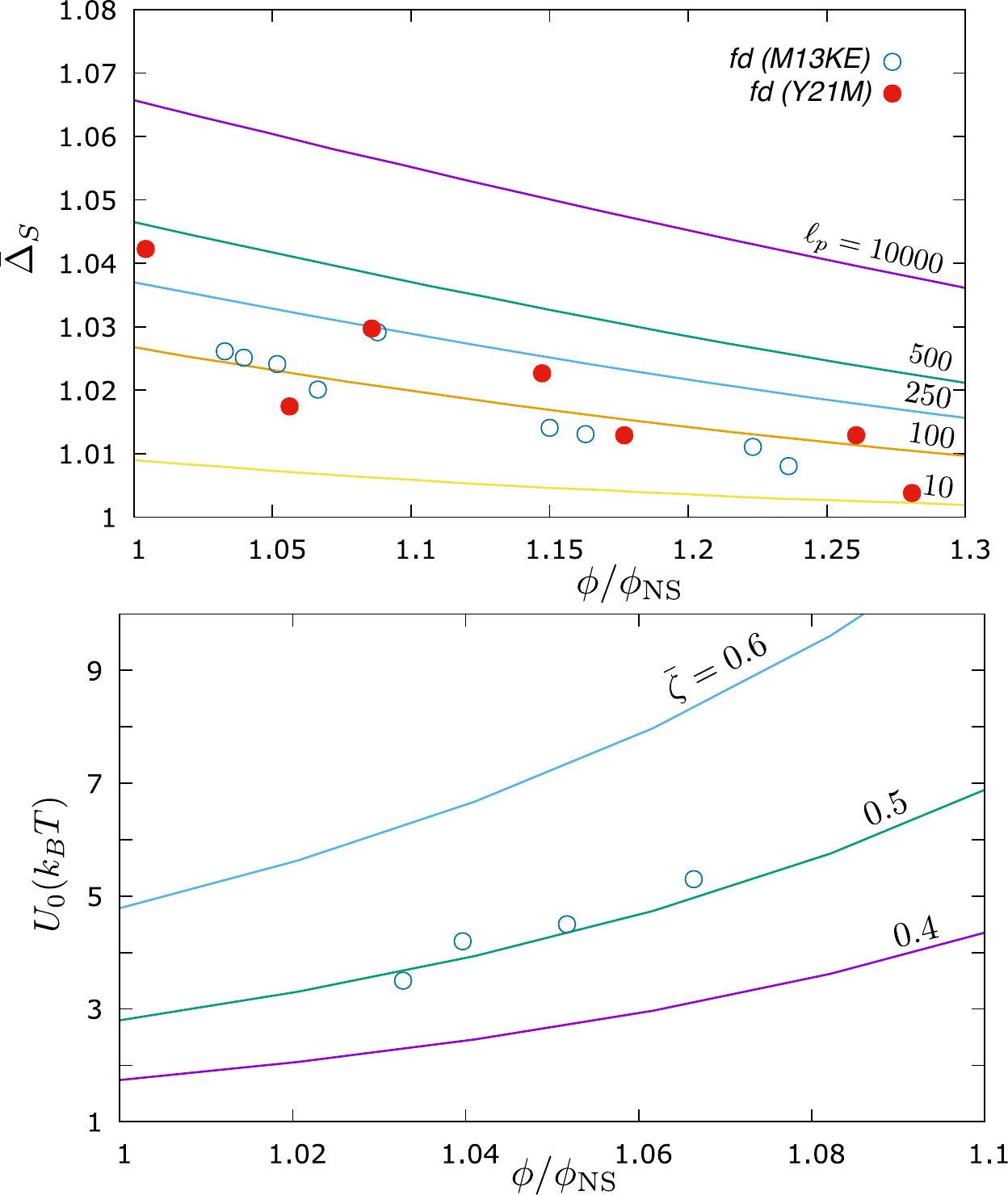}
	\caption{ Theory-experiment comparison based on data for {\em fd} virus rods in a smectic organization. (a) Lamellar spacing versus  rod packing fraction $\phi$ renormalized to the value at nematic-smectic coexistence $\phi_{\rm NS}$. The interdigitation depth is fixed at $\bar{\zeta} = 0.5$. Shown are  results for two different {\em fd} strands: M13KE  ($\ell_{p}/L =3$) and the much stiffer Y21M  ($\ell_{p}/L=10$). (b) Smectic potential for a number of different interdigitation depths $\bar{\zeta} = \zeta/L$ and $\ell_{p} = 250$. Experimental data are based on M13KE ($\ell_{p} = 3$). }
	\label{fdcomp}
\end{figure}

\section{Comparison with experimental results: colloidal versus thermotropic smectics}

We are now in a position to recompute quantities such as the smectic layer spacing, splay and compression moduli and penetration length for stiff rods by considering a large but finite persistence $\ell_{p}/L \gg 1 $ and compare with experimental measurements for {\em fd} rods. Given that the intralamellar nematic order parameter $\alpha_{\perp}$ is no longer strictly quadratic with rod density we must take \eq{splayalpha} rather than \eq{psplay} to compute the splay modulus. The rigid rod results shown in \fig{comp} and \fig{pene}  are recovered simply by taking the limit $\ell_{p}/L \rightarrow \infty$.  The limit of stability  for the smectic phase  at lower rod packing fraction can be gleaned from the nematic-smectic bifurcation \eq{nsbif} which now also depends on the persistence length.  The results (not shown here) are in line with the scenario emerging from a more elaborate theoretical analysis by van der Schoot \cite{van1996nematic}, namely an increase of the nematic-smectic instability density and a simultaneous reduction of the smectic layer spacing with increasing rod flexibility. The reduction of the layer spacing with $\ell_{p}$  is  demonstrated in \fig{fdcomp} where we show a comparison between our model predictions and experimental data for {\em fd} rods in the smectic concentration range. Recently, the penetration elastic length was measured to be   $\lambda    \simeq 0.02 \pm 0.01~\mu$m which, considering the micrometer contour length of filamentous viruses, gives $\lambda/L \approx 0.02 \pm 0.01$ \cite{Repula2018}  which is in outstanding agreement with the predictions showcased in \fig{pene} for rigid rods.  More specifically,  taking an effective persistence length $\ell_{p}/L =100$ and interdigitation depth $\bar{\zeta} = 0.5$ we find $\lambda/L = 0.019 $ in quantitative agreement with the experimental value at least for the particular concentration considered in  experiment ($\phi/\phi_{\rm NS} = 1.36$) \cite{Repula2018}.

Let us now compare our findings with results reported for thermotropic smectics which are stabilized by attractive forces acting between the mesogens rather than through a trade-off between orientation versus volume-exclusion entropy. Taking the lamellar spacing $\Delta_{S}$ as the intrinsic length scale one finds that generically $\lambda \sim \Delta_{S}$. Quoting experimental values from Refs. \cite{ribotta1974quasielastic,Birecki1976,Fromm1987,Lewis1988} we specify $\lambda \sim 1$ to 7 nm with $\Delta_{S} \sim 3$~nm \cite{Davidov1979}. By contrast,  for colloidal smectics composed of {\em fd} rods, $\lambda $ amounts to  only a fraction of the lamellar spacing. The main reason behind this discrepancy is that the typical layer compressibility modulus $B \sim 10^{3-4}$ ${\rm N/m^{2}}$  for colloidal rod smectics is found to be at least three orders of magnitude smaller than the typical value $B \sim 10^{7}$ ${\rm N/m^{2}}$ encountered for thermotropic smectics \cite{ribotta1974quasielastic,Birecki1976,Benzekri1992,Carbone2011}. The splay constant, however, is of similar magnitude for both colloidal and thermotropic smectics. For a typical thermotropic smectic we quote from Refs. \cite{ribotta1974quasielastic,Birecki1976} a splay director diffusion coefficient $K_{1}/\eta_{s} \sim 2 \times 10^{-10}$ ${\rm m^{2}/s}$ which, taking a typical (splay) viscosity $\eta_{s} \sim 10^{-2} $ ${\rm Pa .s}$, leads to a modulus of about $K_{1}  \sim 10 $  ${\rm pN}$. This value is also in agreement with the study of Bradshaw and Raynes \cite{Bradshaw1983}. For colloidal {\em fd} rods, the experimental value for the splay elastic constant is found to be about $2\pm 1~{\rm pN}$ \cite{Repula2018}. This estimate is based on: 1) measurement of the bend modulus $\kappa$ of fluid-like membranes composed of a monolayer of aligned viruses \cite{Barry2010}, which gives after normalization by the lamellar spacing $\Delta_{S}$: $K_1=\kappa /\Delta_{S}\approx 1$~pN; 2) the nematic splay elastic constant evaluated close to the nematic-smectic transition inferred from  Odijk's scaling prediction (Eq.~\ref{K1nem}) for rigid hard rods \cite{odijk1986elastic} $K_1 \approx 3 K_2 \approx 3$~pN, where $K_2$ is the twist elastic constant measured from unwinding the helical mesostructure of a  cholesteric virus suspension under a magnetic field \cite{Dogic2000}. The experimental value is thus found to be at most an order of magnitude smaller than our prediction $K_{1} \sim 10 $   ${\rm pN}$ according to \eq{psplay} taking typical values for {\em fd} ($\ell =100$, $\phi =0.5$, $D =10$ ${\rm nm}$ and $\xi L^{2} =200$). The discrepancy could be attributed to the absence of conformational fluctuations in our treatment of the volume-exclusion entropy  as well as the neglect of electrostatic interactions in our model.  

\section{Conclusion}

Inspired by a recent upsurge in experimental work on colloidal smectics,  we have proposed a microscopic theory that combines concepts from  density functional and cell theory in an effort to predict the thermodynamics and elastic properties of colloidal smectic phases stabilized primarily by entropic forces due to rod volume exclusion. In our study, we focus on predicting the layer compression and bending moduli, which for colloidal smectics have scarcely been addressed thus far, and we quantify those as a function of the rod packing fraction and rod aspect ratio. The characteristic length scale associated with the ratio of the elastic moduli, called the smectic penetration length, amounts to  only a fraction of the layer spacing in contrast to thermotropic smectic materials where the two lengthscales are found to be similar.  
We attribute this to the fact that smectics composed of colloidal rods experience a much smaller layer compressibility ($B$) than typical thermotropic smectics whereas the energy scales associated with layer bending ($K_{1}$) are of comparable magnitude. This illustrates that colloidal smectics have a  mechanical response that is considerably different from their thermotropic counterparts with key implications for their defect topology.  

We further demonstrate that introducing small degree of backbone flexibility, such as present in filamentous virus 
rods, leads to a reduction of the equilibrium lamellar spacing. Our predictions for the spacing, penetration length and layer ``barrier-hopping" potential which governs the single-particle diffusive dynamics along the layer normal, are in quantitative agreement with experimental measurements.  

Since our theory is restricted to hard-core volume exclusions alone, the effect of electrostatic repulsion and other ``soft" interactions as well as the subtle role of  rod flexibility in the volume-exclusion entropy is overlooked here and should be addressed in more elaborate numerical treatments including computer simulations as has been done semi-flexible case \cite{milchev2018nematic, milchev2019smectic, de2017self}. We believe that this issue is particularly relevant for stiff but non-rigid rods with $\ell_{p} >L$ such as filamentous {\em fd} virus discussed here, where backbone flexibility is usually ignored but is  found to play an important role nevertheless in determining  the smectic mesostructure and dynamics through the ``hopping" potential \cite{grelet2014hard,lettinga2007self}.  Although {\em fd} rods are strongly charged and strictly do not interact as hard objects, their phase behavior can be mapped quite efficiently on a hard-core model using a suitable renormalization of the rod dimensions \cite{grelet2014hard}. Similar recipes could be applied to  other smectogenic colloidal particles provided the soft interactions are not too strong and long-ranged. Our predictions will then be instrumental, for instance, in guiding continuum models \cite{xia2021structural, paget2022smectic,paget2023complex} to classify topological defects in a wide range of lyotropic smectic materials.

\begin{acknowledgments}

We acknowledge financial support  from the French National Research Agency (ANR) under  grant ANR-19-CE30-0024 ``ViroLego". We also kindly thank Jaydeep Mandal and Prabal Maiti from Bangalore University (India) for constructive discussions and for sharing their simulation results with us. 

\end{acknowledgments}

\bibliographystyle{unsrt}
\bibliography{refs}

\section*{Appendix: Incipient freezing from a uniform nematic fluid}

In this Appendix we wish to illustrate the wider applicability of the analytical rendering of the excluded volume between elongated colloidal particles proposed in  Section III in relation to phase transitions in systems of rod and disc-shaped colloids.  We  use a simple stability  analysis to predict the nematic-smectic or nematic-columnar transition at the level of bifurcations from a spatially uniform nematic fluid. We apply this to a number of relevant systems, namely  rods and discs in 3D and rods confined to a smectic 
 monolayer.  An overview of the results is given in Table I. Details of the analysis for each specific case can be found in the paragraphs below.

\begin{table*}
\centering
\begin{tabular}{ |p{3cm}||p{3cm}|p{3cm}|p{3cm}|p{3cm}|  }
 \hline
 \multicolumn{5}{|c|}{nematic-solid (NX), nematic-smectic (NS), and nematic-columnar (NC)  freezing of hard cylinders} \\
 \hline
transition & $\phi^{\ast}$ parallel  & $\phi^{\ast}$ freely rotating &  $q^{\ast}$ parallel  & $q^{\ast}$ freely rotating  \\
 \hline
 NX   monolayer rods    & {\bf 0.58}  & {\bf 0.81}  & {\bf  5.13$/D$} & {\bf 5.13$/D$}\\
 NS rods in 3D   & {\bf 0.34} &  {\bf 0.40} &   {\bf 4.49$/L$} & {\bf 4.86$/L$} \\
NC rods in 3D &   0.44  & 0.73 &    5.13$/D$ & 5.13$/D$ \\
NS discs in 3D  & {\bf 0.34} & 0.75 &    {\bf 4.49$/L$} &  4.49$/L$ \\
 NC discs in 3D & 0.44 &  {\bf 0.61} &    5.13$/D$ & {\bf 5.45$/D$} \\  
 \hline
\end{tabular}
\caption{Overview of bulk freezing transitions in terms of the bifurcation packing fraction $\phi^{\ast}$ for  hard cylinders in a  two-dimensional monolayer fluid and in three dimension bulk nematics.   The corresponding wavenumber defining the typical distance between adjacent smectic layers or columns  is indicated by  $q^{\ast}$. The relevant transition values are indicated in bold. }
\end{table*}

The case of a monolayer rod fluid, which is relevant for thin films or strongly confined systems of elongated macromolecules, has received recent attention from density functional theory \cite{oettel2016monolayers}, experiment \cite{klopotek2017monolayers} as well as computer simulation \cite{rajendra2022packing}. In the latter study, for strictly flat monolayers a nematic-solid transition was identified around  a packing fraction $83 \%$ independent of the aspect ratio \cite{prico}.  This value is very close to our prediction ($81 \%$) listed in Table I. In the context of space-filling smectics solidification of a single smectic monolayer can be tentatively connected to the emergence of SmB order. Our results then suggests that the transition from SmA to SmB order occurs at very high packing conditions. It is likely then, however,  that the crossover from SmA to SmB order for strictly hard cylinders is pre-empted by transitions to solid phases that are thermodynamically more stable under these conditions \cite{bolhuis1997tracing,mcgrother1996re,mederos_overview2014}.

For slender rods in a 3D bulk nematic, the results are in good agreement with computer simulation results where the nematic-smectic transition is found to vary only weakly with aspect-ratio \cite{bolhuis1997tracing,mcgrother1996re}. This complies with the ``rule-of-thumb" that a critical packing fraction of a little over 40 \% seems required to stabilize a frozen state in hard body systems, irrespective of their shape \cite{frenkel1989invited,frenkel2015order}.  For strictly parallel rods the transition would happen at a far too low packing fraction, namely $\phi^{\ast} =0.34$. We conclude  that orientational fluctuations play an essential role in steering the transition towards a realistic density range.

Our treatment of the excluded-volume correlations can be further tested by scrutinizing the nematic-columnar transition for thin discs ($L/D \ll 1$). For strictly parallel particles the excluded volume is {\em identical} to the one for elongated cylinders that a second-virial theory would not be able to distinguish between oblate and prolate cylinders. For discs, however,  we know from simulations \cite{frenkel1989invited,veerman1992phase,marechal2011phase} and experiments on clay platelets \cite{van2000liquid,mourad2010lyotropic,davidson2005mineral} that the nematic-columnar transition preempts the nematic-smectic transition  whereas second-virial theory for parallel cylinders predicts that columnar order is {\em always} metastable with respect to the smectic phase \cite{mulder1987density}. While it is known since Onsager's seminal work \cite{onsager1949effects} that second-virial approaches generally fall short in quantitatively describing discotic systems \cite{you2005equation}, we argue that the orientation fluctuations of the discs (ignored in parallel-core models \cite{mederos_overview2014})  also play a key role in favoring columnar over smectic ordering. This is corroborated by our theory (see Table I) in that freely rotating hard discs favor columnar over smectic order, even though the critical packing fraction at the  nematic-columnar transition ($\phi^{\ast} = 0.61$) is overestimated by our theory in line with earlier theoretical findings \cite{wensink2009phase}.

\subsection*{Freezing  of a monolayer rod fluid }

When the intralamellar density exceeds a critical value the monolayer fluid is expected to transition into a solid characterized by in-plane periodic order following some Bravais lattice that we may parameterize by a (combination of) wavevectors ${\bf q}_{\perp}$. Close to the nematic-solid (NX) transition, that we assume to be of a second-order nature \cite{rajendra2022packing,prico},  the one-body density can be written in terms of the reference solution for the fluid state and a periodic density modulation with infinitesimally small amplitude $\epsilon \ll 1$ \cite{mulder1987density,xu1992nematic,allen1993hard}
\beq
\rho( \borpe ; \oma) = \rho_{\perp} \left [ f(\oma) + \epsilon f_{q} (\oma) e^{i {\bf q}_{\perp} \cdot \borpe} \right ],
\eeq
where the orientation distribution $f_{q}(\oma) \neq f(\oma)$ in general depends on the wavevector ${\bf q}_{\perp}$ of the in-plane density modulation.

Next, we linearize the grand potential \eq{fex} for arbitrarily small amplitude $\epsilon$. The result is a bifurcation condition that  identifies the point where spatially non-uniform distributions for the planar one-body density branch off from the fluid solution
\beq
 f_{q} (\oma) = \rho_{\perp} f(\oma) \int d \omb  f_{q}(\omb) \hat{\mathcal{K}}( {\bf q}_{\perp} ). 
\label{rhobif}
\eeq
The key quantity is the kernel $\hat{\mathcal{K}}$  that combines the hard-core volume-exclusion contributions discussed in the main text
\begin{align}
 \rho_{\perp} \hat{\mathcal{K}}( {\bf q} _{\perp})  &=  -8\phi_{\perp}g(\phi_{\perp}) \left \{ \frac{J_{1} (q_{\perp}) }{q_{\perp}} +  \mathcal{F} ({\bf q}_{\perp})  \right \}.
\end{align}
 The first terms denotes the parallel hard-core term while the second term accounts for the orientational fluctuations of the rods and features an explicit coupling between the incipient density modulation and the azimuthal rod angle
\begin{align}
\mathcal{F} ({\bf q}_{\perp})  &=   \frac{\gamma \ell}{3 \pi} \sum_{n=1}^{3}  j_{0} [ \tfrac{q_{\perp} }{2} \ell  ( \theta {\bf a}_{n} \cdot \oma_{\perp}  - \theta^{\prime} {\bf a}_{n} \cdot \omb_{\perp})  ], 
\label{khc_scale}
\end{align}
with $\oma_{\perp} = (\sin \varphi, \cos \varphi , 0 )$.
Here, the density modulation describes, for instance,  a hexagonal Bravais lattice  with primitive vectors ${\bf a}_{1} =  (\sqrt{3}, 1)/2$ and ${\bf a}_{2} =  (-\sqrt{3}, 1)/2$ and ${\bf a}_{3} = (0,1)$. Other crystal symmetries may be probed likewise \cite{baus1987statistical}.

Symmetry-breaking solutions deviating from a uniform fluid towards a solid state may be probed from the condition \eq{rhobif}. In the simplest case, we assume that the orientational distribution  does {\em not respond} to the incipient density modulation and obeys a Gaussian form \eq{fgauss}. This amounts to replacing in \eq{rhobif}
\beq
f_{q} (\oma) = f_{G} (\oma). 
\eeq
The bifurcation condition \eq{rhobif} is then equivalent to a divergence of the static structure factor indicating a loss of local stability of the monolayer fluid 
\begin{align}
& S^{-1}( q_{\perp} ) =   1 + 8\phi_{\perp} g(\phi_{\perp}) \left \{ \frac{J_{1} (q_{\perp}) }{q_{\perp}}  + \langle \langle \mathcal{F} ({\bf q}_{\perp} ; ) \rangle \rangle  \right \}  =0.
\label{sinfrozen}
\end{align}
The brackets denote a double Gaussian average defined in the main text. The bifurcation solution of \eq{sinfrozen} then corresponds to the lowest (real-positive) value of $\phi_{\perp}^{\ast}$ with the associated wavenumber indicated by $q_{\perp}^{\ast}$.
The averages of the interrod angles are known analytically in the asymptotic limit and are given in \eq{gammav}. 
This is not the case, however, for the angular average  of $\mathcal{F}$ where the polar and azimuthal angles are strongly convoluted with the wavevector ${\bf q}_{\perp}$. To make headway we assume that the polar angles of the rods  are `frozen' at a small value $\theta = \theta^{\prime} = \theta_{c} \ll 1$ while fluctuations in the azimuthal angles remain \textit{a priori} unrestricted. Next we substitute the asymptotic expression $ \gamma \approx ( \theta^{2} + \theta^{\prime 2} - 2 \theta \theta^{\prime} \cos \Delta \varphi )^{1/2} \sim 2^{1/2} \theta_{c} \sqrt{1- \cos \Delta \varphi}$ with $\Delta \varphi = \varphi - \varphi^{\prime}$.  Technically this amounts to replacing the Gaussian orientational probability by an even  simpler factorized form
\beq
f_{G}(\oma ) \sim  \frac{\delta (\theta - \theta_{c})}{\sin \theta_{c}} \frac{1}{2 \pi}.
\label{ffact}
\eeq
Since the azimuthal fluctuations are presumed unaffected by the  density modulation, the three lattice vector contributions ${\bf a}_{n}$ are equal. Using the result $\tfrac{2^{1/2}}{2 \pi}\int_{0}^{2\pi} dx \sqrt{1- \cos x}  = 4/\pi$, the problem reduces to a single azimuthal average
\begin{align}
& \langle \langle \mathcal{F} ({\bf q}_{\perp}) \rangle \rangle  \sim  \frac{4}{\pi^{2}} (\ell \theta_{c}) \int_{0}^{2 \pi} \frac{d\varphi}{2 \pi}      j_{0} [ \tfrac{q_{\perp} }{2} (\ell \theta_{c}) \cos \varphi  ]. 
\label{fsimple}
\end{align}
We may link the constrained polar angle $\theta_{c}$ to the nematic order parameter $\alpha_{\perp}$ and rod packing fraction  $\phi_{\perp}$ via  
\beq
\ell \theta_{c} \sim \ell \langle \theta^{2} \rangle_{f_{G}}^{1/2} \sim (2 \ell^{2} / \alpha_{\perp})^{1/2}. 
\eeq
We invoke the  quadratic relationship with the packing fraction $\phi_{\perp}$  \eq{alpha} and find
\beq
\ell \theta_{c} \sim \left ( \frac{\pi}{2} \right )^{1/2} \frac{1}{\phi_{\perp} g(\phi_{\perp})},
\eeq
independent from the rod aspect ratio $\ell$.
A trivial reference case is a system of perfectly parallel hard cylinders previously analyzed by Mulder \cite{mulder1987density}.  We set $\theta_{c} =0$ and $f(\oma) = \delta ( \oma)$ and find that the structure factor \eq{sinfrozen} has a pole at $\phi_{\perp} \approx 0.58$ and $q_{\perp} \approx 5.13$. The presence of orientational fluctuations expressed by \eq{fsimple}] leads to a much higher transition values namely $\phi_{\perp} \approx 0.81$ while leaving the critical wavenumber unchanged ($q_{\perp} \approx 0.513$). The results have been tabulated in Table I in the main text.

\subsection*{Smectic versus columnar freezing of hard rods}

In order to further test our theory, we consider the much more widely studied system  of rigid hard cylinders in 3D \cite{frenkel1988thermodynamic, allen1993hard, mcgrother1996re, bolhuis1997tracing, mederos_overview2014}. When exceeding a certain critical packing fraction these systems are known to form a smectic-A phase emerging from a nematic fluid. In our approach, the hard-body interaction between the cylinders can be approximated by a parallel hard-core contribution supplemented with a fluctuation term that depends on the rod orientations. Ignoring correlations between end-caps which should be negligible for sufficiently long rods $\ell = L/D \gg 1$, we find that FT of the excluded volume reads as follows
\begin{align}
& \hat{\mathcal{K}}({\bf q} )  \approx  -2 \pi LD^{2} j_{0}(q_{\parallel}L)\frac{J_{1}(q_{\perp}D)}{\tfrac{1}{2} q_{\perp} D} \nonumber \\ 
& - 2 L^{2} D | \sin \gamma | j_{0}(\tfrac{L}{2} {\bf q} \cdot \oma ) j_{0}(\tfrac{L}{2} {\bf q} \cdot \omb ) + \mathcal{O}(D^{3}).
\label{knem}
\end{align}
Assuming the principal director to point along the $z$-axis of the lab frame we write $q_{\parallel} = {\bf q} \cdot \hat{\bf z}$ and $q_{\perp} = {\bf q} \cdot \hat{\bf x} = {\bf q} \cdot \hat{\bf y}$. Similar to the monolayer case we may probe bifurcations from the uniform nematic fluid towards, for instance, a smectic structure  for which ${\bf q} \cdot \oma = q_{\parallel} \cos \theta $ and $q_{\perp}  =0$. Then, the nematic-smectic (NS) bifurcation in the asymptotic limit of strong alignment follows from 
\beq
S_{NS}^{-1}(q_{\parallel}) \sim 1 +  8\phi g(\phi) \left [ j_{0}(q_{\parallel} L)   + \tfrac{1}{\pi} \ell \langle \langle  \gamma \rangle \rangle  j_{0}^{2}(\tfrac{q_{\parallel}L}{2})  \right ]  =0. 
\label{nsbif}
\eeq 
 Using the Gaussian approximation we write for the double averaged angle
\beq
\ell \langle \langle \gamma  \rangle \rangle \sim \frac{\pi}{2 g(\phi) \phi}. 
\eeq
The solution is $\phi^{\ast} = 0.403$ and $q_{\parallel}^{\ast}L = 4.86$  {\em independent} of the rod aspect ratio $\ell$. 

To finalize our analysis we also explore the possibility of a nematic-columnar transition for rods  in which case a density modulation develops across the plane transverse to the nematic director  ${\bf q} = (q_{\perp} , 0 , 0 )$. In the asymptotic limit of near-parallel rods we find $D{\bf q} \cdot \oma \sim \mathcal{O} (q_{\perp}D \theta ) \sim 0 $ so that the divergence criterion for the structure factor reads
\beq
S_{NC}^{-1}(q_{\perp}) \sim 1 +  8\phi g(\phi) \left [ \frac{J_{1} (q_{\perp} D)}{\tfrac{1}{2} q_{\perp} D}   + \tfrac{1}{\pi} \ell \langle \langle  \gamma \rangle \rangle  \right ]  =0, 
\eeq 
with solution $\phi^{\ast} = 0.74$ and $q_{\perp}^{\ast}D = 5.14$. This  demonstrates that the nematic-smectic transition strongly pre-empts the columnar phase as is well known from computer simulation \cite{frenkel1988thermodynamic,bolhuis1997tracing,mcgrother1996re} and experiment \cite{meyer1990ordered,dogic1997smectic}.

\subsection*{Smectic versus columnar freezing of hard discs}

We finish our analysis by addressing freezing instabilities in fluids of thin cylindrical discs with a large diameter-to-thickness ratio $D/L \gg 1$.
 The orientational fluctuation contribution to the excluded volume for thin discs is much more complicated than the one for rods and has been computed by one of us \cite{wensink2014generalized}
\begin{align}
\hat{\mathcal{K}}(\oma, \omb)  \approx & -2 \pi LD^{2} j_{0}(q_{\parallel}L)\frac{J_{1}(q_{\perp}D)}{\tfrac{1}{2} q_{\perp} D} \nonumber \\ 
& -\frac{\pi}{4} D^{3} | \sin \gamma | \left ( A_{1} \frac{J_{1}(\bar{q}_{2}) }{\tfrac{1}{2} \bar{q}_{2}} +  A_{2} \frac{J_{1}(\bar{q}_{1}) }{\tfrac{1}{2} \bar{q}_{1}} \right ) \nonumber \\ 
& + \mathcal{O}(DL^{2}), 
\end{align}
with $\bar{q}_{n} = \sqrt{(\tfrac{D}{2} {\bf q} \cdot \hat{\bf v})^{2}  + (\tfrac{D}{2} {\bf q} \cdot \hat{\bf w}_{n})^{2}  }  $ and
\begin{align}
A_{n} &= \tfrac{1}{2}\int_{-1}^{1} dt \cos ( \tfrac{D}{2} t {\bf q} \cdot \hat{\bf w}_{1} ) \cos (\tfrac{D}{2} \sqrt{1-t^{2}} {\bf q} \cdot \hat{\bf v} ) \nonumber \\
& \approx J_{0}(\bar{q}_{n}), 
\end{align}
with $\hat{\bf v} = (\oma \times \omb) / | \sin \gamma | $ and  $\hat{\bf w}_{n} = {\boldsymbol \omega}_{n} \times \hat{\bf v}  $ two auxiliary unit vectors associated with the particle frame of each disc. It is easily verified that the zero-wavenumber limit $(\bar{q}_{n} = 0) $ of the second contribution above yields $- \tfrac{\pi}{2} D^{3} | \sin \gamma |$ which corresponds to (minus) the excluded volume between two infinitely thin hard discs of diameter $D$. Taking the asymptotic limit of small polar angles and considering only density modulations in the $xy-$ plane perpendicular to the nematic director ($q_{\parallel}=0$) we find that the orientation-dependent arguments of the Bessel functions can be expressed as
\beq
\bar{q}  \sim  \pm q_{\perp} D \left ( \frac{\theta^{\prime}  - \theta \cos \Delta \varphi }{2\gamma}  \right ).
\eeq
The sign is irrelevant here given that both Bessel functions above are even functions.
The next step is to pre-average the term between brackets over the disc orientations which in the Gaussian approximations gives a mere constant of $\mathcal{O}(1)$, namely 
$\langle \langle ( \theta^{\prime}  - \theta \cos \Delta \varphi) /2\gamma \rangle \rangle  = c_{0}  \approx 0.3$.  Taking all this into account we find that the divergence of the structure factor at the nematic-columnar (NC) transition can be established from the following  simple condition
\begin{align}
& S_{NC}^{-1}(q_{\perp}) \sim 1 +  8\phi g(\phi) \nonumber \\ 
& \times \left [ \frac{J_{1} (q_{\perp} D)}{\tfrac{1}{2} q_{\perp} D}   +  \tfrac{1}{2} \delta \langle  \langle  \gamma \rangle \rangle \frac{ J_{0}( c_{0} q_{\perp} D)J_{1}( c_{0} q_{\perp} D) }{c_{0} q_{\perp} D} \right ]  =0, 
\end{align} 
where $\delta = D/L \gg 1$ defines the aspect-ratio of the disc. Applying the Gaussian scaling arguments invoked for rods to the case of discs we find  
\beq
\delta \langle \langle \gamma  \rangle \rangle \sim \frac{2}{g(\phi) \phi}, 
\eeq
which renders the results universal and independent of the disc aspect ratio. The bifurcation criterion is easily resolved numerically and yields $\phi^{\ast} = 0.61$ and $q_{\parallel}^{\ast} L = 5.45$. Similarly, we may probe the nematic-smectic transition from taking ${\bf q} = q_{\parallel} \hat{\bf z}$. Then, in the asymptotic limit we infer that $\bar{q}  \sim \mathcal{O}(  q_{\parallel} L \theta)$ is small and can be set to zero in good approximation. The condition for the nematic-smectic transition for discs then reads
\begin{align}
& S_{NS}^{-1}(q_{\parallel}) \sim 1 +  8\phi g(\phi) \left [ j_{0}(q_{\parallel} L)    +  \tfrac{1}{2} \delta \langle  \langle  \gamma \rangle \rangle  \right ]  =0, 
\end{align}
which has a solution  $\phi^{\ast} = 0.75$ and $q_{\parallel}^{\ast} L = 4.49$. The results are summarized in Table I.

\end{document}